\documentclass[12pt,notitlepage]{revtex4}
\usepackage{graphicx}

\usepackage{amsmath}
\usepackage{amsfonts}
\usepackage{amssymb}
\usepackage{slashbox}
\usepackage{epstopdf}

\font\msytw=msbm9 scaled\magstep1 \font\msytww=msbm7
scaled\magstep1 \font\msytwww=msbm5 scaled\magstep1

\let\a=\alpha \let\b=\beta  \let\g=\gamma  \let\d=\delta \let\e=\varepsilon
  \let\h=\eta   \let\th=\theta  \let\l=\lambda
\let\m=\mu    \let\n=\nu    \let\x=\xi     \let\p=\pi    \let\r=\rho
\let\s=\sigma \let\t=\tau   \let\f=\varphi \let\c=\chi
   \let\o=\omega
\let\G=\Gamma \let\D=\Delta  \let\L=\Lambda 
         
\let\O=\Omega 

\def\PPP{{\cal P}}\def\EE{{\cal E}} \def\VV{{\cal V}}
 \def\WW{{\cal W}}
\def\TT{{\cal T}} \def\BBB{{\cal B}}
\def\RR{{\cal R}}\def\LL{{\cal L}}  
\def\DD{{\cal D}} \def\SS{{\cal S}}

 \def\pp{{\bf p}}\def\Ve{{\bf e}}
 \def\xx{{\bf x}} \def\yy{{\bf y}} \def\zz{{\bf z}}
\def\kk{{\bf k}}
\def\PP{{\bf P}}\def\dd{{\boldsymbol{\d}}}

\def\nn{\nonumber}

\def\RRR{\hbox{\msytw R}} \def\rrrr{\hbox{\msytww R}}

\def\NNN{\hbox{\msytw N}} 
 \def\ZZZ{\hbox{\msytw Z}}
 \def\zzz{\hbox{\msytwww Z}}

\def\\{\hfill\break}
\def\={:=}
\let\io=\infty
\let\0=\noindent
\def\media#1{{\langle#1\rangle}}
\let\dpr=\partial

\def\const{{\rm const}}
\def\tende#1{\,\vtop{\ialign{##\crcr\rightarrowfill\crcr\noalign{\kern-1pt
    \nointerlineskip} \hskip3.pt${\scriptstyle #1}$\hskip3.pt\crcr}}\,}
\def\otto{\,{\kern-1.truept\leftarrow\kern-5.truept\to\kern-1.truept}\,}

\def\to{\rightarrow}

\def\qed{\hfill\raise1pt\hbox{\vrule height5pt width5pt depth0pt}}

\def\ul#1{{\underline#1}}
\def\lis{\overline}
\def\V#1{{\bf#1}}
\def\be{\begin{equation}}
\def\ee{\end{equation}}
\def\bea{\begin{eqnarray}}
\def\eea{\end{eqnarray}}
\def\nn{\nonumber}
\def\pref#1{(\ref{#1})}

\def\Tr{\mathrm{Tr}}
\def\eu{\mathrm{e}}

\newtheorem{lemma}{Lemma}[section]

\begin{document}
\title{Universality of conductivity in interacting graphene}
\author{A. Giuliani} \affiliation{Universit\`a di
Roma Tre, L.go S. L. Murialdo 1, 00146 Roma - Italy}
\author{V. Mastropietro}
\affiliation{Universit\`a di Roma Tor Vergata, V.le della Ricerca
Scientifica, 00133 Roma - Italy}
\author{M. Porta}
\affiliation{ETH, Wolfgang Pauli Strasse 27, 8093 Z\"urich - Switzerland}
\begin{abstract} The Hubbard model on the honeycomb lattice
describes charge carriers in graphene with short range interactions. While
the interaction modifies several physical quantities, like the
value of the Fermi velocity or the wave function renormalization,
the a.c. conductivity has a universal value independent of the
microscopic details of the model: there are no interaction corrections,
provided that the interaction is weak enough and that the system is at 
half filling. We give a rigorous proof of this fact, based 
on exact Ward Identities and on constructive Renormalization Group methods.
\end{abstract}

\maketitle

\renewcommand{\thesection}{\arabic{section}}

\section{Introduction and main results}\label{sec1}
\setcounter{equation}{0}
\renewcommand{\theequation}{\ref{sec1}.\arabic{equation}}


The effects of interactions in quantum many body theory at low
temperatures pose notoriously difficult problems; in certain cases
the physical properties of the system are radically changed with respect to the
non interacting situation, while in other cases, like in the so-called 
Fermi liquids, a simple modification or {\it
renormalization} of the physical quantities is expected. There is,
however, a very small group of phenomena which are {\it universal};
the physical quantities appear to be protected from any {\it
renormalization} due to the interactions and their values do not dependent 
on the details of the model, but rather upon fundamental constants. 
A celebrated example is provided by the
Quantum Hall Effect, in which the value of the plateaus only depend on the 
von Klitzing constant $h/e^2$ and not on the material parameters;
universality appears to be related to topological invariance
\cite{T,AA} or to the presence of Ward Identities \cite{IM}.
Other examples of universal phenomena come from the physics of
superconductivity, in which the magnetic quantum $h/e$ plays an
important role.

In recent times, evidence for universality has been observed in the
conductivity of {\it graphene}, a one atom thick layer of
graphite. Its electronic properties can be well described in terms of
a tight-binding model of electrons hopping from one site to a
neighboring one of a honeycomb lattice, but often this model is {\it
approximated} by an effective one expressed in terms of massless
Dirac fermions in the two-dimensional continuum \cite{S}. 
Recent optical measurements \cite{N1}
show that at half-filling and small temperatures, if the frequency
is in a range well inside the temperature and the band-width, the
a.c. conductivity is essentially constant and equal, up to a few
percent, to $\s_0=\frac{e^2}{h}\frac{\pi}{2}$. Such value only
depends on the von Klitzing constant and not on the material
parameters, like the Fermi velocity; it is apparently {\it
universal}, at least inside the experimental precision.

Is graphene a.c. conductivity truly universal? Theoretically, the
computation of the conductivity in the {\it absence of
interactions} gives exactly the value
$\s_0=\frac{e^2}{h}\frac{\pi}{2}$, both in the idealized 
{\it Dirac description} \cite{A} and even in the more realistic tight binding 
model \cite{SPG}. However, since truly universal
phenomena are quite rare in condensed matter, it is important to
understand whether this apparently {\it universal value} is just
an artifact of the idealized description in terms of
non-interacting fermions or rather it is a robust property still
valid in the presence of electron-electron interactions, which are
certainly present and expected to play a role in real graphene.
Such a question has been studied in the physical literature, but
contradictory results have been found, see \cite{SS1,M1, H0}. The
reason for this is that the Dirac approximation, which works well
for the free gas, is not very accurate in the presence of
interactions: the corrections to the conductivity are expressed
by logarithmically {\it divergent integrals}. One can argue that such
divergence is spurious, just an artifact of the Dirac approximation, and
that a regularization must be adopted to cure it
(in the tight binding model, the lattice provides a
natural cut-off); however, the results appear to be regularization-dependent 
and no unique predictions can be drawn.

In this paper we consider the Hubbard model on the honeycomb
lattice, as a model of monolayer graphene with screened
interactions, and we prove that the a.c. conductivity has the
universal value $\s_0=\frac{e^2}{h}\frac{\pi}{2}$ even in presence
of interactions: {\it the interaction corrections to the
conductivity are vanishing}, provided that the interaction is weak
enough and the system is in the half filled band case. Remarkably, 
the presence of the lattice and its symmetries is essential to get the result.
The idea of the proof is based on the two main ingredients: (i)
{\it exact lattice} Ward Identities (WI) relating the
current-current, vertex and 2-point functions; (ii) the fact that
the interaction-dependent corrections to the Fourier transform of
the current-current correlations are {\it differentiable} with
continuous derivative (in contrast, the free part is continuous
and not differentiable at zero frequency).

The paper is organized in the following way. In Section \ref{sec1a} we describe
the model, in Section \ref{sec1ab} we derive the Ward Identities, in 
Section \ref{sec1ac} (and Appendix \ref{appA}) 
we perform the computation of the conductivity in the non
interacting case, and in Section \ref{sec1ad} we present the proof of our
main result, under some regularity assumptions on the
current-current correlations. The proof of these regularity properties, 
the full description of the expansion for the current-current correlations and 
vertex functions, as well as the proof of convergence of this expansion, 
is given in Section \ref{sec1.2b} (Appendix \ref{secB} and \ref{secD2} collect 
the proof of some symmetry properties extensively used in Section 
\ref{sec1.2b}). 

\subsection{The model and the observables}\label{sec1a}

We consider electrons on a two-dimensional honeycomb lattice interacting via a
local Hubbard interaction, as a model describing the charge carriers in
graphene. Its ground state properties at half-filling, including
the asymptotic behavior of the correlations at large distances, have
already been analyzed in \cite{GM,GM1}. In this subsection we recall
the definition of the model and introduce some of the key observables
(density and current), which will allow us to define the conductivity, i.e.,
the quantity of main interest in this paper.\\

{\it The fermionic fields and the Hamiltonian.}
Let $\L=\{n_1\vec l_1+n_2\vec l_2\ :\  n_{1}, n_{2}=0,\ldots,L-1\}$
be a periodic triangular lattice of period $L$, with basis vectors:
${\vec l_1}=\frac12(3,\sqrt{3})$, ${\vec l_2}=\frac12(3,-\sqrt{3})$.
Let us denote by $\L_A=\L$ and $\L_B=\L+\vec\d_i$ the $A$- and $B$-sublattices
of the honeycomb lattice, with $\vec \d_i$ the n.n. vectors defined as:
\be {\vec \d_1}=(1,0)\;,\quad {\vec \d_2}=\frac12
(-1,\sqrt{3})\;,\quad{\vec \d_3}=\frac12(-1,-\sqrt{3})\;.\label{1.1}\ee
If $\vec A$ and $\vec B$ are two arbitrary constant vectors, we introduce
creation and annihilation fermionic operators for electrons sitting at the
sites of the $A$- and $B$- sublattices with spin index $\s=\uparrow\downarrow$
as
\bea && a^\pm_{\vec x,\s}=L^{-2}\sum_{\vec k\in\BBB_L}e^{\pm i\vec k
(\vec x-\vec A)}\hat a^\pm_{\vec k,\s}\;,\qquad\vec x\in\L_A\;,\label{1.e}\\
&& b^\pm_{\vec x,\s}=L^{-2}\sum_{\vec k\in\BBB_L}e^{\pm i\vec k
(\vec x-\vec B)}\hat b^\pm_{\vec k,\s}\;,\qquad\vec x\in\L_B\;,\label{1.f}\eea
where $\BBB_L=\{\vec k=n_1\vec G_1/L+ n_2\vec G_2/L\,:\, 0\le
n_i<L\}$, with $\vec G_{1,2}=\frac{2\p}3(1,\pm\sqrt3)$, is the first
Brillouin zone; note that in the thermodynamic limit
$L^{-2}\sum_{\vec k\in\BBB_L}\to|\BBB|^{-1} \int_{\BBB} d\vec k$,
with $\BBB= \{\vec
k=\x_1\vec G_1+\x_2 \vec G_2\,:\, \x_i\in[0,1)\}$ and
$|\BBB| =8\pi^2/(3\sqrt{3})$. The operators $a^\pm_{\vec x,\s},
b^\pm_{\vec x,\s}$ satisfy the canonical anticommutation rules and are periodic
over $\L$; their Fourier transforms are normalized in such a way that, if
$\vec k,\vec k'$ are both in the first Brillouin zone:
\be \{\hat a^\e_{\vec k,\s},\hat a^{\e'}_{\vec k',\s'}\}=L^2\d_{\vec k,\vec k'}
\d_{\e,-\e'}\d_{\s,\s'}\;,\qquad \{\hat b^\e_{\vec k,\s},
\hat b^{\e'}_{\vec k',\s'}\}=L^2\d_{\vec k,\vec k'}
\d_{\e,-\e'}\d_{\s,\s'}\;.\label{1.4}\ee
Moreover, they are quasi-periodic over the first Brillouin zone:
\be a^\pm_{\vec k+\vec G_i,\s}=e^{\pm i\vec G_i\cdot\vec A}
a^\pm_{\vec k,\tau}\;,\qquad b^\pm_{\vec k+\vec G_i,\tau}=e^{\pm i\vec G_i
(\vec B-\vec \d_j)} b^\pm_{\vec k,\tau}\;,\label{1.4b}\ee
where $e^{i\vec G_i\vec\d_j}=e^{i 2\pi/3}$, for all values of $i,j$.
The phases $\vec A$ and $\vec B$ are arbitrary, the freedom in their
choice corresponding to the freedom in the choice
of the origins of the two sublattices $\L_A$ and $\L_B$ (this
symmetry is sometimes referred
to as {\it Berry-gauge} invariance). A convenient choice for $\vec A$ and
$\vec B$, which makes the fields $\hat a^\pm_{\vec k,\s},
\hat b^\pm_{\vec k,\s}$ periodic over the reciprocal lattice $\L^*$, is
$\vec A=\vec 0$ and $\vec B=\vec\d_1$, which reads:
\be a^\pm_{\vec x,\s}=L^{-2}\sum_{\vec k\in\BBB_L}e^{\pm i\vec k \vec x}
\hat a^\pm_{\vec k,\s}\;,\qquad
b^\pm_{\vec x+\vec\d_1,\s}=L^{-2}\sum_{\vec k\in\BBB_L}e^{\pm i\vec k\vec x}
\hat b^\pm_{\vec k,\s}\;,\qquad\vec x\in\L_A\;.\label{1.fg}\ee
This is the choice made in \cite{GM,GM1}, which will be used throughout this
paper, too.

The grand-canonical Hamiltonian of the two-dimensional Hubbard model
on the honeycomb lattice at half-filling is $H_\L=H^0_\L(t)+U
V_\L$, where $H^0_\L(t)$ is the free Hamiltonian, describing nearest
neighbor hopping ($t\in\RRR$ is the hopping parameter):
\be H^0_\L(t)=H^0_\L(\{t_{\vec x,j}\})\big|_{t_{\vec x,j}\equiv t}\;,\qquad
H^0_\L(\{t_{\vec x,j}\})= -\sum_{\substack{\vec x\in\L_A \\
j=1,2,3}}\sum_{\s=\uparrow\downarrow}(t_{\vec x,j}
a^{+}_{\vec x,\s} b^{-}_{\vec x +\vec
\d_j,\s}+t_{\vec x,j}^*b^{+}_{\vec x +\vec
\d_j,\s}a^{-}_{\vec x,\s} )\;,\label{1}\ee
and $V_\L$ is the local Hubbard interaction:
\be V_\L=\sum_{\substack{\vec x\in \L_A\\
i=1,2,3}}\Big(a^+_{\vec x,\uparrow} a^-_{\vec
x,\uparrow}-\frac12\Big)\Big(a^+_{\vec x,\downarrow} a^-_{\vec
x,\downarrow}-\frac12\Big)+\sum_{\substack{\vec x\in \L_B\\
i=1,2,3}}\Big(b^+_{\vec x,\uparrow} b^-_{\vec
x,\uparrow}-\frac12\Big)\Big(b^+_{\vec x,\downarrow} b^-_{\vec
x,\downarrow}-\frac12\Big)\;.\label{2}\ee
\vskip.2truecm
{\it The current and density operators.}
The {\it current} is defined as usual via the {\it Peierls
substitution}, by modifying  the hopping parameter along the bond
$(\vec x,\vec x+\vec\d_j)$ as
\be t\to t_{\vec x,j}(\vec A)=t\,e^{ie\int_0^1\vec A(\vec
x+s\vec\d_j)\cdot\vec\d_j\,ds}\;, \ee
where the constant $e$ appearing at exponent is the electric charge
and $\vec A(\vec x)\in\RRR^2$ is a periodic field on ${\cal S}_\L= \{\vec
x=L\x_1\vec l_1+L\x_2 \vec l_2\,:\, \x_i\in[0,1)\}$. Its Fourier
transform is defined as $\vec A(\vec x)=|\SS_\L|^{-1}\sum_{\vec
p\in\DD_\L} \vec A_{\vec p}e^{-i\vec p\vec x}$, where
$|\SS_\L|=\frac{3\sqrt3}2L^2$ and $\DD_\L=\{\vec p=n_1\vec G_1/L+
n_2\vec G_2/L\,:\, n_i\in\ZZZ\}$; note that in the thermodynamic
limit $|\SS_\L|^{-1}\sum_{ \vec
p\in\DD_\L}\to(2\p)^{-2}\int_{\rrrr^2} d\vec p$. If we denote by
$H_\L(\vec A)= H_\L^0(\{t_{\vec x,j}(\vec A)\})+UV_\L$ the modified
Hamiltonian with the new hopping parameters, the lattice current
is defined as $\vec J_{\vec p}^{\,(A)}=-|\SS_\L|\,\dpr H_\L(\vec A)/\dpr \vec
A_{\vec p}$, which gives, at first order in $\vec A$,
\be \vec J^{\,(A)}_{\vec p}=\vec J_{\vec p}+\frac1{|\SS_\L|}
\sum_{\vec q\in\DD_\L} \hat \D_{\vec p,\vec q}
\vec A_{\vec q}\;,\label{1.3}\ee
where, if $\h^j_{\vec p}=\frac{1-e^{-i\vec p\vec \d_j}}{i\vec p\vec \d_j}$,
\bea \vec J_{\vec p}&=& iet\,\frac1{L^2}
\sum_{\substack{\vec k\in\BBB_L\\ \s,j}}\,\vec \d_j \h^j_{\vec p}
\big(\hat a^+_{\vec k+\vec p,\s}\hat b^-_{\vec k,\s}
e^{-i\vec k(\vec
\d_j-\vec\d_1)}-\hat b^+_{\vec k+\vec p,\s}\hat a^-_{\vec k,\s}
e^{+i(\vec k+\vec p)(\vec\d_j-\vec\d_1)}\big)=\nn\\
&=& iet\,\sum_{\substack{\vec x\in\L\\ \s,j}}\,e^{-i\vec p\vec x}
\vec \d_j\h^j_{\vec p}\big(a^+_{\vec x,\s}b^-_{\vec x+\vec \d_j,\s}-
b^+_{\vec x+\vec \d_j,\s} a^-_{\vec x,\s}\big)=:ev_0
\sum_{\substack{\vec x\in\L\\ j=1,2,3}}\,e^{-i\vec p\vec x}\vec \d_j
\h^j_{\vec p} J_{\vec x}^j\label{1.4a}\eea
is the {\it paramagnetic current} (in the last rewriting, $v_0=3t/2$ is the
{\it free Fermi velocity} and $J^j_{\vec x}=(2i/3)\big(a^+_{\vec x,\s}
b^-_{\vec x+\vec \d_j,\s}- b^+_{\vec x+\vec \d_j,\s} a^-_{\vec x,\s}\big)$ are
the {\it bond currents}) and
\be\big[\,\hat \D_{\vec p,\vec q}\big]_{lm}=
\sum_{\substack{\vec x\in\L\\ j=1,2,3}}e^{-i(\vec p+\vec q)\vec x}(\vec \d_j)_l
(\vec \d_j)_m \h^j_{\vec p}\h^j_{\vec q}\D_{\vec x,j}\;,\label{1.5}\ee
with $\D_{\vec x,j}=-e^2 t\sum_{\s}(a^+_{\vec x,\s}b^-_{\vec x+\vec\d_j,\s}+
b^+_{\vec x+\vec\d_j,\s}a^-_{\vec x,\s})$, is the {\it diamagnetic tensor}.
Similarly, the {\it density} operator is
defined by coupling the local density to an external field, i.e., by adding to
the Hamiltonian a local chemical potential term of the form
\be M_\L(\m)=-\sum_{\vec x\in\L_A}\sum_{\s=\uparrow\downarrow}\m(\vec x)
a^+_{\vec x,\s}a^-_{\vec x,\s}-\sum_{\vec x\in\L_B}\sum_{\s=\uparrow\downarrow}
\m(\vec x)b^+_{\vec x,\s}b^-_{\vec x,\s}\;,\label{1.5d}\ee
where $\m(\vec x)=|\SS_\L|^{-1}\sum_{\vec
p\in\DD_\L} \m_{\vec p}e^{-i\vec p\vec x}$
is a periodic field on ${\cal S}_\L$.
If we denote by $H_\L(\m)=H^0_\L(t)+UV_\L+M_\L(\m)$ the modified Hamiltonian in
the presence of the local chemical potential, the lattice density is defined as
$\hat\r_{\vec p}=-|\SS_\L|\,\dpr H_\L(\m)/\dpr \m_{\vec p}$, which gives
\bea
\hat\rho_{\vec p}&=&\frac1{L^2}\sum_{\substack{\vec k\in\BBB_L\\ \s=\uparrow
\downarrow}}\big(\hat a^+_{\vec k+\vec p,\s}\hat a^-_{\vec k,\s}+
e^{-i\vec p\vec\d_1}\hat b^+_{\vec k+\vec p,\s}\hat b^-_{\vec k,\s}\big)=
\label{1.5dd}\\&=&
\sum_{\substack{\vec x\in\L_A\\ \s=\uparrow\downarrow}}e^{-i\vec p\vec x}
a^+_{\vec x,\s}a^-_{\vec x,\s}+\sum_{\substack{\vec x\in\L_B\\
\s=\uparrow\downarrow}}e^{-i\vec p\vec x}b^+_{\vec x,\s} b^-_{\vec x,\s}=:
\sum_{\vec x\in\L_A}e^{-i\vec p\vec x}
\r^A_{\vec x}+\sum_{\vec x\in\L_B}e^{-i\vec p\vec x}
\r^B_{\vec x}\;.\nn\eea
It will be convenient for the incoming discussion to
think the two components of the paramagnetic current $\hat J_{\vec p,l}$,
$l=1,2$, as the spatial components of a ``space-time" three-components vector
$\hat J_{\vec p,\m}$, $\m=0,1,2$, with $\hat J_{\vec p,0}=e\hat\r_{\vec p}$.
In the following, it will also convenient to introduce the reduced current
$\vec\jmath_{\vec p}$, related to the paramagnetic current by
\be \vec J_{\vec p}=v_0\vec\jmath_{\vec p}\;,\label{1.jsmall}\ee
with $v_0=3t/2$ the {\it free Fermi velocity}.
\\

{\it Schwinger functions and response functions.}
The thermal state of the system at inverse temperature $\b>0$, associated to
the density matrix $e^{-\b H_\L}$, can be characterized in terms of
{\it Schwinger functions} and {\it response functions}. The
Schwinger functions are the analytic continuation to imaginary time of the
off-diagonal elements of the reduced density matrices; the response functions
give us informations about the reaction of the system to a diversity of
external probes within the linear response regime. They are defined as follows.
Let $\psi_{\vec x,\s}^\pm=(a^\pm_{\vec x,\s}, b^\pm_{\vec x+\vec\d_1,\s})$,
let $O^{(i)}_{\vec x_i}$, $i=1,\ldots,n$, be local monomials in the
$\psi_{\vec x,\s}^\pm$ operators and let us denote by
$O^{(i)}_{\xx_i}=e^{x_{i,0}H_\L}O^{(i)}_{\vec x_i}e^{-x_{i,0}H_\L}$
the corresponding imaginary-time evolved operators;
here $\xx_i=(x_{i,0},\vec x_i)$ and $x_{i,0}\in[0,\b)$ is the imaginary time.
The average of a product of local operators in the thermal state of the
system at inverse temperature $\b>0$ is defined as
\be \media{O^{(1)}_{\xx_1}\cdots O^{(n)}_{\xx_n}}_{\b,L}=
\frac{\Tr\{e^{-\b H_\L}{\bf T}(O^{(1)}_{\xx_1}\cdots O^{(n)}_{\xx_n})\}}
{\Tr\{e^{e^{-\b H_\L}}\}}\;,\label{1.ave}\ee
where ${\bf T}$ is the operator of fermionic time ordering, acting on a
product of fermionic fields as:
\be {\bf T}(\psi^{\e_1}_{\xx_1,\s_1,\r_1}\cdots
\psi^{\e_n}_{\xx_n,\s_n,\r_n})= (-1)^\p
\psi^{\e_{\p(1)}}_{\xx_{\p(1)},\s_{\p(1)},\r_{\p(1)}}\cdots
\psi^{\e_{\p(n)}}_{
\xx_{\p(n)},\s_{\p(n)},\r_{\p(n)}}\label{1.4aa}\ee
where $\e_i\in\{+,-\}$, $\s_i\in\{\uparrow,\downarrow\}$, $\r_i\in\{1,2\}$,
$\xx_i\in[0,\b)\times\L$, $\psi^\pm_{\xx,\s,1}=a^\pm_{\xx,\s}$ and, if
${\boldsymbol{\d}}_i=(0,\vec\d_i)$, $\psi^\pm_{\xx,\s,2}=
b^\pm_{\xx+{\boldsymbol{\d}}_1,\s}$. Moreover,
$\p$ is a permutation of $\{1,\ldots,n\}$, chosen in such a
way that $x_{\p(1)0}\ge\cdots\ge x_{\p(n)0}$, and $(-1)^\p$ is its
sign. [If some of the time coordinates are
equal each other, the arbitrariness of the definition is solved by
ordering each set of operators with the same time coordinate so
that creation operators precede the annihilation operators.]
Finally, we denote by $\media{O^{(1)}_{\xx_1};\cdots ;O^{(n)}_{\xx_n}}_{\b,L}$
the corresponding {\it truncated expectations}. We shall also use the notation
$\media{\cdot}_\b=\lim_{L\to\infty}\media{\cdot}_{\b,L}$ and
$\media{\cdot}=\lim_{L\to\infty}\media{\cdot}_{\b}$.

Choosing the local operators $O^{(i)}_{\xx_i}$ in Eq.(\ref{1.ave}) simply as
monomials in the fermionic fields, we get the
Schwinger functions of order $n$:
\be S_n^{\b,L}(\xx_1,\e_1,\s_1;\ldots;\xx_n,\e_n,\s_n)_{\r_1,\ldots,\r_n}
=\media{\psi^{\e_1}_{\xx_1,\s_1,\r_1}\cdots \psi^{\e_n}_{\xx_n,\s_n,\r_n}
}_{\b,L}\;.\label{1.3a}\ee
Choosing the operators $O^{(i)}_{\xx_i}$ as suitable combinations of the
current and density operators, we get the current-current, density-density and
current-density response functions:
\bea\hat K^{\b,L}_{\m\n}(\pp):=\frac1{\b L^2}\int_{-\b/2}^{\b/2}\!\!\!dx_0\!
\int_{-\b/2}^{\b/2}\!\!\! d y_0\, e^{-i p_0 (x_0-y_0)} \media{J_{(x_0,\vec p),
\m};J_{(y_0,- \vec p),\n}}_{\b,L}\;,\label{1.jj}\eea
where $\pp=(p_0,\vec p)\in\frac{2\p}{\b}\ZZZ\times\DD_\L$ and
$J_{(x_0,\vec p),\m}=e^{x_0 H_\L}\hat J_{\vec p,\m}e^{-x_0 H_\L}$. An
important role will be also played in the following by the two- and
three-points functions:
\be \hat S^{\b,L}(\kk):=\frac1{\b L^2}
\int_{(\b,L)}\!\!\!\!d\xx \int_{(\b,L)}\!\!\!\!d\yy\,e^{+i \kk(\xx-\yy)}
S_2^{\b,L}(\xx,-,\s;\yy,+,\s)\;,\label{1.4s2}\ee
\be \hat G_{2,1;\m}^{\b,L}(\kk,\pp):=\frac1{\b L^2}\int_{-\b/2}^{\b/2}
\!\!\!\!\!\!dx_0 \int_{-\b/2}^{\b/2}\!\!\!\!\!\!dy_0
\int_{-\b/2}^{\b/2}\!\!\!\!\!\!dz_0 \,e^{+i k_0(x_0-y_0)+ip_0(x_0-z_0)}
\media{J_{(z_0,\vec p),\m};\psi^-_{(x_0,\vec k+\vec p),\s
}\psi^+_{(y_0,\vec k),\s}}_{\!\b,L}\label{1.3p}\ee
where $\int_{(\b,L)} d\xx$ is a shorthand for
$\int_{-\b/2}^{\b/2} dx_0\sum_{\vec x\in\L}$ and $\psi^\pm_{(x_0,\vec k),\s}=
e^{x_0 H_\L}\hat \psi_{\vec k,\s}e^{-x_0 H_\L}$, with
$\hat\psi^\pm_{\vec k,\s}=(\hat a^\pm_{\vec k,\s}, \hat b^\pm_{\vec k,\s})$.
Here and in the following, we will exploit whenever possible
the vectorial structure of $\psi$ and the
tensorial structure of products of $\psi$'s; to this purpose,
we will think of $\psi^-$ as a column vector and
$\psi^+$ as a row vector, so that, e.g.,
$\psi^+_{\xx,\s}\psi^-_{\yy,\s}$ will be naturally thought as a scalar, while
$\psi^-_{\xx,\s}\psi^+_{\yy,\s}$ will be thought as a $2\times2$ matrix.
In particular, both $\hat S^{\b,L}(\kk)$ and
$\hat G_{2,1;\m}^{\b,L}(\kk,\pp)$ (for any fixed choice of
$\m\in\{0,1,2\}$) can be thought as $2\times2$ matrices.\\

{\it Conductivity.} The ac {\it conductivity}  in units such that $\hslash =1$
is related to the current-current correlations via Kubo formula \cite{SPG},
which reads, for all $2\p\b^{-1}\ZZZ\ni p_0\neq 0$ and $l,m\in\{1,2\}$:
\be \s^{\b}_{lm}(p_0)= -\frac{2}{3\sqrt3}\frac1{p_0}
\Big[\hat K^{\b}_{lm}(p_0,0)+\D^\b_{lm}\Big]\;,\label{1.sigma}\ee
where $\hat K^{\b}_{lm}(\pp) = \lim_{L\to+\infty} \hat K^{\b,L}_{lm}(\pp)$ and
\be \D^\b_{lm}=\lim_{L\to\infty}\frac1{L^2}
\sum_{\substack{\vec x\in\L\\ j=1,2,3}}
(\vec \d_j)_l(\vec \d_j)_m \media{\D_{\vec x,j}}_{\b,L}\label{1.5az}\ee
is the diamagnetic contribution (see Eq.(\ref{1.5})) and the factor
$2/(3\sqrt3)$ appearing in
Eq.(\ref{1.sigma}) must be understood as the inverse of the area of the unit
cell of the hexagonal lattice.

The main goal of this paper is to compute $\s^{\b}_{lm}(p_0)$
in the zero temperature and zero frequency limit (taking the
limits in a suitable order, so to make contact with experiments on the optical
conductivity of graphene), i.e., to compute the so-called {\it universal
optical conductivity}:
\be \s_{lm}:=\lim_{p_0\to 0^+}\lim_{\b\to\infty}\s^{\b}_{lm}(p_0)\;.
\label{1.uc}\ee
A key role in its computation will be played by Ward
Identities, which show that the quantities introduced above are not
independent; on the contrary, they are related by exact identities, which
we now describe.

\subsection{Conservation laws and Ward Identities}\label{sec1ab}

By definition of $\r_{(x_0,\vec p)}=e^{x_0 H_\L}\hat\r_{\vec p}e^{-x_0 H_\L}$,
we have:
\be  \partial_{x_0} \r_{(x_0,\vec p)}=[H_\L, \r_{(x_0,\vec
p)}]\;.\label{1.9a}\ee
Computing explicitly the r.h.s. of this equation and using, in particular,
the fact that $[V_\L,\r_{\vec p}]=0$, we find:
\be \dpr_{x_0}\r_{(x_0,\vec p)}=t\sum_{\substack{\vec x\in\L\\ \s,j}}
e^{-i\vec p\vec x}(1-e^{-i\vec p\vec \d_j})(a^+_{\xx,\s}b^-_{\xx+\dd_j,\s}-
b^+_{\xx+\dd_j,\s}a^-_{\xx,\s})\;.\label{1.wi2}\ee
Comparing the r.h.s. of this equation with the definition of the paramagnetic
current, we recognize that Eq.(\ref{1.wi2}) can be rewritten as
a {\it continuity equation}:
\be -ie\dpr_{x_0}\r_{(x_0,\vec p)}+i{\vec p}\cdot\vec J_{(x_0,\vec p)}=0\;.
\label{1.wi3}\ee
%

Using the continuity equation Eq.(\ref{1.wi3}), we can easily derive
an exact identity relating three- and two-point functions. In fact,
by the definition of $\hat G_{2,1;0}^{\b,L}(\kk,\pp)$, see
Eq.(\ref{1.3p}), and integrating by parts, we find:
\be ip_0\hat G_{2,1;0}^{\b,L}(\kk,\pp)=
 \frac1{L^2}\int_{-\b/2}^{\b/2}\!\!\!\!\!\!dy_0
\int_{-\b/2}^{\b/2}\!\!\!\!\!\!dz_0 \,e^{-i k_0y_0-ip_0z_0}
\dpr_{z_0}\!\media{J_{(z_0,\vec p),0};\psi^-_{(0,\vec k+\vec p),\s
}\psi^+_{(y_0,\vec k),\s}}_{\!\b,L}\qquad\hfill\quad\label{1.wi30}\ee
The derivative with respect to $z_0$ can act either on $J_{(z_0,\vec p),0}$,
in which case we use the continuity equation, or on the Heaviside step
functions involved in the definition of time-ordering operator ${\bf T}$.
After some straightforward algebra, and using the fact that
\be [\hat\r_{\vec p},\hat\psi^-_{\vec k+\vec p,\s}]=-M(\vec p)
\hat\psi^-_{\vec k,\s}\;,\qquad
[\hat\r_{\vec p},\hat\psi^+_{\vec k,\s}]=
\psi^+_{\vec k+\vec p,\s}M(\vec p)\;,\label{1.wi4}\ee
with
\be M(\vec p)=\begin{pmatrix}1&0\\ 0&
e^{-i\vec p\vec \d_1}\end{pmatrix}\;,\label{1.wi4a}\ee
we end up with
\be -ip_0\hat G_{2,1;0}^{\b,L}(\kk,\pp)+p_1\hat G_{2,1;1}^{\b,L}(\kk,\pp)+
p_2\hat G_{2,1;2}^{\b,L}(\kk,\pp)=-e\hat S^{\b,L}(\kk+\pp)M(\vec p)+
eM(\vec p)\hat S^{\b,L}(\kk)\;.\label{1.14} \ee
Proceeding in the same way, we also find:
\bea && -ip_0\hat K^{\b,L}_{0,0}(\pp)+p_1\hat K^{\b,L}_{1,0}(\pp)+
p_2\hat K^{\b,L}_{2,0}(\pp)=0\;,\label{1.wi21}\\
&& -ip_0\hat K^{\b,L}_{0,m}(\pp)+p_1\hat K^{\b,L}_{1,m}(\pp)+p_2
\hat K^{\b,L}_{2,m}(\pp)
=-\frac1{L^2}\Big[\vec p\cdot \media{\hat\D_{\vec p,-\vec p}}_{\b,L}\Big]_m
\;,\label{1.wi22}
\eea
where $m=1,2$, and we used the fact that
\be e[\hat\r_{\vec p},\vec J_{-\vec p}]=
\sum_{\substack{\vec x\in\L\\ j=1,2,3}}(\vec p\cdot\vec\d_j)\,\vec\d_j
|\h^j_{\vec p}|^2\D_{\vec x,j}=\vec p\cdot \hat\D_{\vec p,-\vec p}\;.
\label{1.wi5}\ee
The term in the r.h.s. of Eq.(\ref{1.wi22}) is known as {\it Schwinger term}.

Note that from Eq.(\ref{1.wi22}), setting, e.g., $p_2=0$, we find that,
for $i=1,2$,
\be \hat K_{1i}(p_0,p_1,0)+\lim_{\b,L\to\infty}\frac1{L^2}
\media{\big[\hat\D_{(p_1,0),(-p_1,0)}\big]_{1i}}_{\b,L}=
i\frac{p_0}{p_1}\hat K_{0i}(p_0,p_1,0)\;,\label{1.wi6}\ee
provided the limits of these functions as $\b,L\to\infty$ exist.

Now, {\it if we knew
that the correlations in Eq.(\ref{1.wi6}) were continuous and continuosly
differentiable at $\pp=\V0$}, from the latter equation we would
conclude that $\lim_{\pp\to \V0}[\hat K_{1i}(\pp)+\D_{1i}]=0$ (with
$\D_{lm}=\lim_{\b\to\infty}\D^\b_{lm}$, see Eq.(\ref{1.5az})); moreover, 
continuous differentiability together with the symmetry properties of our model 
would imply that (with a proof analogous to the one of item (ii) in Proposition $1$ below) 
$\lim_{\pp\to \V0}\dpr_{p_0}\hat K_{1i}(\pp)=0$.
Comparing these equations with the definition of the universal optical
conductivity, see Eqs.(\ref{1.sigma})-(\ref{1.uc}), then
we would be tempted to conclude that $\s_{1i}=0$ (and a similar argument would
imply that $\s_{2i}=0$). {\it This is in contrast with the explicit
computation of
the conductivity in the non-interacting case}, which will be discussed below.
The solution to this apparent paradox is that $\hat K_{lm}(\pp)$ is not
continuosly differentiable at $\pp=\V0$. In fact, as it will turn out from
the following discussion, the regularity properties of the Fourier transform
of the current-current correlations play a crucial role in the physical
properties of the conductivity.

\subsection{Properties of the non interacting model}\label{sec1ac}

In the absence of interactions, that is for $U=0$, the two-points function
defined in Eq.(\ref{1.4s2}) reads (see \cite{GM,GM1}):
\be\hat S^{\b,L}(\kk)\Big|_{U=0}=\frac{1}
{k_0^2 + v_0^2 |\O(\vec k)|^2}\begin{pmatrix}  i k_0 & -v_0\O^{*}(\vec k)
\\ -v_0\O(\vec k) & i k_0 \end{pmatrix}=:S_0(\kk)\;, \label{11}\ee
where $v_0=\frac32 t$ and $\O(\vec k)=\frac23\sum_{j=1}^3
e^{i\vec k(\vec\d_j-\vec\d_1)}$. The {\it complex dispersion relation}
$\O(\vec k)$ vanishes only at the two {\it
Fermi points}
\be \vec p_{F}^{\ \pm}=(\frac{2\pi}{3},\pm\frac{2\pi}{3\sqrt{3}})\;,\label{1.7}\ee
close to which it behaves as follows: $\O(\vec p_F^\pm+\vec k')=i k_1'\pm k_2'
+O(|\vec k'|^2)$. The Schwinger functions of higher order can be
explictly computed in terms of $S_0(\kk)$ via the fermionic Wick rule.

Also the universal optical conductivity can be computed explicitly,
see Appendix \ref{appA}, and turns out to be equal to (restoring the presence
of the dimensional constant $\hslash=h/(2\p)$ in the result):
\be \s_{ij}\big|_{U=0}=
\frac2{3\sqrt{3}}\, \frac{2e^2v_0^2}{\hslash}\lim_{p_0\to 0^+}
\int\frac{d k_0}{2\p}\, \int_{\BBB}\frac{d\vec k}{|\BBB|}{\rm
Tr}\Big\{\frac{S_0(\kk+(p_0,\vec 0))-S_0(\kk)}{p_0} \G_i(\vec k,\vec
0)S_0(\kk)\G_j(\vec k,\vec 0)\Big\}\;,\label{1.cond1}\ee
where
\be \vec \G(\vec k,\vec p)=\frac23\sum_{j=1}^3 \vec\d_j
\begin{pmatrix} 0 & i e^{-i\vec k(\vec \d_j-\vec\d_1)}\\
-i e^{+i(\vec k+\vec p)(\vec\d_j-\vec\d_1)}&0\end{pmatrix}\;.\label{1.cond2}\ee
Note that the r.h.s. of Eq.(\ref{1.cond1}) depends on $v_0$. Moreover,
the integral is not uniformly convergent in $p_0$ as $p_0\to 0$;
in particular, it is well known that one cannot exchange the limit with the
integral \cite{Z}. An explicit computation, see Appendix \ref{appA},
yields
\be  \s_{ij}\big|_{U=0}=\frac{e^2}{h}\frac{\p}{2}\d_{ij}\;,\label{1.cond3}\ee
a value that, remarkably, does not depend on $v_0$.
It is also remarkable that the same value of the conductivity is found
in the so-called {\it Dirac approximation}, that is by replacing $S_0(\kk)$
in Eq.(\ref{1.cond3}) with its linear approximation around the Fermi points,
in the presence of an ultraviolet cutoff, i.e.,
\be \sum_{\o=\pm}\c(|\vec k-\vec p_F^\o|\le \e)\hat
G^{v_0}_\o(k_0,\vec k-\vec p_F^\o)\;,\qquad\hat G^{v_0}_\o(\kk)=
\begin{pmatrix}-ik_0& -v_0(-ik_1+\o k_2)\\ -v_0(ik_1+\o k_2) & -ik_0
\end{pmatrix}^{\!-1}\;,\ee
where $\e$ is a small positive number. Therefore, $\s_{ij}\big|_{U=0}$
does not depend on the lattice parameters and, in this sense,
we can say that the free conductivity is {\it universal}.

\subsection{Universality of the conductivity}\label{sec1ad}

Let us now discuss the effects of the Hubbard interaction. We will first recall
the results of \cite{GM,GM1}, concerning the thermodynamic functions of the
model, the Schwinger functions and, in particular, the
two- and three-points functions. Next, we will state our results concerning
the interacting conductivity.

In Theorem 1 of \cite{GM}, we proved that, if $U$ is small enough
(uniformly in the system size and in the temperature), then the specific
free energy $f_\b(U)$ and the finite temperature Schwinger
functions in the thermodynamic limit are analytic functions of $U$,
uniformly in $\b$ as $\b\to\io$, and so are the specific ground state energy
$e(U)$ and the zero temperature Schwinger functions in the thermodynamic limit.
In particular, we proved that the Fourier transform of
the zero-temperature two-points function in the thermodynamic limit
is singular only
at the Fermi points $\kk=\pp_F^{\pm}=(0,\vec p_F^{\,\pm})$
and, close to the singularities, if $\o=\pm$, it can be written as
\be\hat S(\pp_F^\o+\kk')=
\frac1{Z}\begin{pmatrix}-i k_0 & -v_F\O^*(\vec p_F^{\,\o}+\vec k')\\ -v_F
\O(\vec p_F^{\,\o}+\vec k') & -ik_0\end{pmatrix}^{\!-1}
\Big(\openone + R(\kk')\Big)\;,\label{1.9}\ee
where $Z$ and $v_F$ are two analytic functions of $U$, analytically close
to their unperturbed values,
\be Z=1+ O(U^2)\;,\qquad\qquad v_F=v_0+O(U^2)\;.\label{1.10}\ee
Moreover the matrix $R(\kk')$ satisfies $||R(\kk')||\le C |\kk'|^\vartheta$
for some constants $C, \vartheta>0$ and $|\kk'|$ small enough.
In \cite{GM1} we also announced the following result, whose detailed proof
will be given below.\\

{\bf Theorem 2.} {\it There exists $U_0>0$ such that, the three-point
function Eq.(\ref{1.3p}) is uniformly analytic in $\b,L$ and uniformly
convergent to an analytic function as $\b,L\to\infty$, for all
$\kk\neq\pp_F^\pm$, $\pp\neq \V0$ and for $(\min_{\o}|\kk-\pp_F^\o|)$,
$|\pp|\!\cdot\!(\min_\o|\kk-\pp_F^{\o}|)^{-1}$ sufficiently small.
Moreover, the vertex function at the
Fermi points (in the thermodynamic and zero-temperature limits) is equal to:
\be \lim_{\kk\to\pp_F^\o}\lim_{\pp\to\V0}\,
[\hat S(\kk+\pp)]^{-1}\hat G_{2,1;\m}(\kk,\pp)[\hat
S(\kk)]^{-1}=eZ_\m\G_\m(\vec p_F^\o,\vec 0)\;,\label{thm2}\ee
where $\G_i(\vec k,\vec p)$, $i=1,2$, were defined in Eq.(\ref{1.cond2}),
$\G_0(\vec k,\vec p)=1$ and
$Z_\m=Z_\m(U)$, $\m=0,1,2$, are analytic functions of $U$.}\\

Note that the existence of the limits in Eq.(\ref{thm2}) is part of the
statement of the theorem. An important consequence of Theorem 2
(also announced in \cite{GM1}) is obtained by combining its result with
the WI Eq.(\ref{1.14}): in fact, using Eqs.(\ref{1.9})-(\ref{thm2}) into
Eq.(\ref{1.14}) we find that the vertex functions are related to the
wave function renormalization $Z$ and to the Fermi velocity $v_F$ by simple
identities:
\be Z_0=Z\;,\qquad Z_1=Z_2=v_F Z\;.\label{thm2.1}\ee
These relations can be proven as follows: take the limits in the l.h.s. of
Eq.(\ref{thm2}) with $\kk=\pp_F^\o+k'\Ve_\m$ and $\pp=p'\Ve_\m$,
where $\Ve_0=(1,0,0)$, $\Ve_1=(0,1,0)$ and $\Ve_2=(0,0,1)$;
using Eq.(\ref{1.14}) and Eq.(\ref{1.9}) we can rewrite the l.h.s. of
Eq.(\ref{thm2}) (after having taken $p'\to 0$) as:
\bea&& -e(i)^{\d_{\m,0}}\lim_{k'\to 0} [\hat S(\pp_F^\o+k'\Ve_\m)]^{-1}
\dpr_{\m}\hat S(\pp_F^\o+k'\Ve_\m) [\hat S(\pp_F^\o+k'\Ve_\m)]^{-1}=\nn\\
&&=e(i)^{\d_{\m,0}}\lim_{k'\to 0}\dpr_{\m}[\hat S(\pp_F^\o+k'\Ve_\m)]^{-1}
\label{1.100}\eea
that, using again Eq.(\ref{1.9}), is equal to $eZ(v_F)^{1-\d_{\m,0}}
\G_\m(\vec p_F^\o,\vec 0)$.\\

To summarize, at half filling, weak local electron-electron interactions
do not change the infrared behavior of correlations: they ``just'' change the
values of some physical parameters, namely the wave function renormalization
$Z$, the Fermi velocity $v_F$ and the vertex functions $Z_\m$; the latter are
related to $Z$ and $v_F$ in a simple way, thanks to WIs.
The next natural question
we would like to answer to is how is the conductivity changed by the presence
of interactions. Since the infrared behavior of the interacting correlations
is the same as the non-interacting one, it is natural to expect that the
interacting conductivity remains finite in the zero-temperature and
zero-frequency limit; what is apriori unclear is whether the zero-frequency
conductivity remains universal in any reasonable sense, in analogy with the
universal behavior of the free conductivity. Quite remarkably,
we can prove that the interacting conductivity is universal in a very strong
sense: namely, we prove that $\s_{lm}$ is not only independent of
the details of the lattice, but it is also exactly independent of $U$.\\
\vskip.1cm {\bf Theorem 3.} {\it There exists a constant $U_0>0$ such that,
for $|U|\le U_0$ and any fixed $p_0$ (non vanishing and sufficiently small),
$\s^{\b}_{lm}(p_0)$ is analytic in $U$
uniformly in $\b$ as $\b\to\infty$ and uniformly convergent
to an analytic function of $U$ as $\b\to\infty$. Moreover,
\be \s_{lm}=
\lim_{p_0\to 0^+}\lim_{\b\to\io}\s^{\b}_{lm}(p_0)=
\frac{e^2}{h}\frac{\pi}{2}
\d_{lm}\;.\label{1.7b}\ee }
Note that the limit $\b\to\io$ is taken before the limit $p_0\to
0^+$. In other words, the theorem says that the interaction
corrections to the conductivity are negligible at frequencies
$\b^{-1}\ll p_0\ll t$, in agreement with experiments on the optical
conductivity \cite{N1}.
The above results says that all the interaction corrections to the
conductivity cancel out exactly, even if the Fermi velocity and
the wave function are renormalized by the interaction.\\

The proof of Theorem 3 is based on two
main ingredients: (i) the use of the exact Ward Identities
Eqs.(\ref{1.wi21})-(\ref{1.wi22}); (ii) the
fact that the interaction-dependent corrections to the Fourier
transform of the current-current correlations are {\it
differentiable} with continuous derivative (in contrast, the free part is
continuous and not differentiable at zero frequency). The main technical
point of this paper is to control the regularity properties
of the interaction corrections to the conductivity, which are summarized in the
following proposition.\\

{\bf Proposition 1.} {\it There exists $U_0>0$ such that, if $|U|\le U_0$,
then the current-current function $\hat K^{\b,L}_{lm}(\pp)$
is analytic in $U$, uniformly in $\b,L$, for all sufficiently small $\pp\neq \V0$
and it is uniformly convergent
as $\b,L$ to the function $\hat K_{lm}(\pp)=
\lim_{\b\to\infty}\lim_{L\to\infty}\hat K^{\b,L}_{lm}(\pp)$,
which is also analytic in $U$ for all sufficiently $\pp\neq \V0$. The latter
function satisfies the following properties:
\begin{enumerate}
\item $\hat K_{lm}(\pp)$ is continuous for all sufficiently small $\pp\in\RRR\times
\BBB$ (in particular at $\pp=\V0$) and continuously differentiable
for all sufficiently small $\pp\neq\V0$.
\item $\hat K_{lm}(\pp)$ can be decomposed as:
\be \hat K_{lm}(\pp)= \frac{Z_l Z_m}{Z^2}
\media{\hat\jmath_{\pp,l};\hat\jmath_{-\pp,m}}^{0,v_F}+R_{\m\n}(\pp)\;,
\label{1.jjr}\ee
where:\begin{enumerate}
\item[(i)] if  $\media{\cdot}^{0,v_F}_{\b,L}$ the average with respect to the
density matrix
$e^{-\b H^0_\L(\frac23v_F)}$ associated to the non-interacting Hamiltonian
with Fermi velocity $v_F$, then
$$\media{\hat\jmath_{\pp,l};\hat\jmath_{-\pp,m}}^{0,v_F}=\lim_{\b,L\to\infty}
\frac1{\b L^2}
\int_{-\b/2}^{\b/2}\!\!\!dx_0\int_{-\b/2}^{\b/2}\!\!\!dy_0\,
e^{-i p_0 (x_0-y_0)} \media{j_{(x_0,\vec p),l};
j_{(y_0,-\vec p),m}}^{0,v_F}_{\b,L}\;,$$ with
$j_{(x_0,\vec p),l}=e^{x_0H_\L}j_{\vec p,l}e^{-x_0H_\L}$ and
$\vec\jmath_{\vec p}$
the reduced current defined in Eq.(\ref{1.jsmall});
\item[(ii)]  $R(\pp)$ is continuously differentiable for all sufficiently small $\pp\in \RRR\times
\BBB$ (in particular at $\pp=\V0$); moreover, $R(p_0,\vec 0)=R(-p_0,\vec 0)$.
\end{enumerate}
\end{enumerate}
}

One can immediately realize that Theorem 3 is a simple corollary of
Proposition 1 and of the WIs Eqs.(\ref{1.wi21})-(\ref{1.wi22}). In fact,
from Eq.(\ref{1.wi22}) computed at $\pp=(0,p_1,0)$, we find
\be\hat K_{1i}(0,p_1,0)+\lim_{\b,L\to\infty}\frac1{L^2}
\media{\big[\hat\D_{(p_1,0),(-p_1,0)}\big]_{1i}}_{\b,L}=0\label{1.jj2}\ee
that implies, using the continuity of $\hat K_{lm}(\pp)$
at $\pp=\V0$ stated in Proposition 1,
\be\lim_{\pp\to\V0}\hat K_{1i}(\pp)=-\D_{1i}\label{1.jj3}\ee
and a similar argument shows that
\be\lim_{\pp\to\V0}\hat K_{lm}(\pp)=-\D_{lm}\label{1.jj4}\ee
for all $l,m\in\{1,2\}$. Therefore, using again the continuity at $\pp=\V0$
of the current-current function and the definition of conductivity, we can
rewrite
\be \s_{lm}=-\frac2{3\sqrt3}\lim_{p_0\to 0^+}\frac1{p_0}
\Big[\hat K_{lm}(p_0,\vec 0)-\hat K_{lm}(\V0)\Big]\;.\label{1.jj5}\ee
We now use the decomposition Eq.(\ref{1.jjr}) to rewrite the latter
equation as
\bea \s_{lm}&=&-\frac2{3\sqrt3}\lim_{p_0\to 0^+}\frac1{p_0}
\frac{Z_lZ_m}{Z^2}
\Big[\media{\hat\jmath_{(p_0,\vec 0),l};\hat\jmath_{-(p_0,\vec 0),m}}^{0,v_F}-
\media{\hat\jmath_{\V0,l};\hat\jmath_{\V0,m}}^{0,v_F}\Big]\nn\\
&&-\frac2{3\sqrt3}\lim_{p_0\to 0^+}\frac1{p_0}
\Big[R_{lm}(p_0,\vec 0)-R_{lm}(\V0)\Big]\;.\label{1.jj6}\eea
Now, using the identity $Z_i=Zv_F$, see Eq.(\ref{thm2.1}), we conclude that the
limit in the first line reduces to the computation of the free
conductivity Eq.(\ref{1.cond1}) with $v_0$ replaced by $v_F$: however,
since the result does not depend on the Fermi velocity, from the first
line we simply get $(e^2/h)(\p/2)\d_{lm}$.
On the other hand, the limit in the second line,
using the continuous differentiability of $R(\pp)$, reduces to
$-\frac2{3\sqrt3}\dpr_0R(\V0)$, which is zero, simply because $R(\pp)$ is even
in $\pp$.

This concludes the proof of Theorem 3, once that Proposition 1 is given.
The rest of the paper will be devoted to proofs of Theorem 2 and Proposition 1.

\section{Regularity of the current-current correlations}\label{sec1.2b}
\setcounter{equation}{0}
\renewcommand{\theequation}{\ref{sec1.2b}.\arabic{equation}}

In this section we prove Theorem 2 and Proposition 1. We will use an extension
of the method discussed in \cite{GM}. We will assume the reader familiar with
the proof in \cite{GM} and we will only describe in detail the new aspects
of the construction, as compared to the one in \cite{GM}. Still, we will try to
be as self-consistent as possible, possibly giving reference to a well-defined
section of \cite{GM} for the few technical aspects that will not be fully
reproduced here.

\subsection{Grassman Integral representation for the correlation functions}

The goal is to prove analyticity in $U$ and regularity in $\kk,\pp$ of the
three-point and current-current functions. We remind the reader that the proof
of analyticity and the control of regularity of the Schwinger functions
(in particular, of the two-points function) has already appeared in \cite{GM},
see in particular Section 3.4 of \cite{GM}.
The starting point of our construction is a representation of the generating
function for correlations in terms of a Grassmann functional integral,
completely analogous to the one used in \cite{GM} to write the generating
function for the Schwinger functions. The Grassmann functional integral we
are intersted in is defined as follows.

Let $M\in\NNN$ and $\c_0(t)$ a smooth compact support function
that is $1$ for $t\le a_0$ and $0$ for $t\ge 2 a_0$, with $a_0$ a constant
that can be chosen equal to, e.g., $1/3$ (see the condition
on $a_0$ appearing after Eq.(3.41) of \cite{GM} and read it for $\g=2$). Let
$\BBB_{\b,L}^*=\big\{2\p\b^{-1}(\ZZZ+\frac12)\cap
\{k_0\,:\,\c_0(2^{-M}|k_0|)>0\}\big\}
\times\BBB_L$. We consider the
finite Grassmann algebra generated by the Grassmannian variables
$\{\hat\Psi^\pm_{\kk,\s,\r}\}_{ \kk \in
\BBB_{\b,L}^*}^{\s=\uparrow\downarrow,\ \r=1,2}$ and a Grassmann
integration $\int
\big[\prod_{\kk\in\BBB^*_{\b,L}}\prod_{\s=\uparrow\downarrow}^{\r=1,2}
d\hat\Psi_{\kk,\s,\r}^+ d\hat\Psi_{\kk,\s,\r}^-\big]$ defined as
the linear operator on the Grassmann algebra such that, given a
monomial $Q( \hat\Psi^-, \hat\Psi^+)$ in the variables
$\hat\Psi^\pm_{\kk,\s,\r}$, its action on $Q(\hat\Psi^-,
\hat\Psi^+)$ is $0$ except in the case $Q(\hat\Psi^-,
\hat\Psi^+)=\prod_{\kk\in\BBB_{\b,L}^*}
\prod_{\s=\uparrow\downarrow}^{\r=1,2} \hat\Psi^-_{\kk,\s,\r}
\hat\Psi^+_{\kk,\s,\r}$, up to a permutation of the variables. In
this case the value of the integral is determined, by using the
anticommuting properties of the variables, by the condition
\be \int
\Big[\prod_{\kk\in\BBB_{\b,L}^*}\prod_{\s=\uparrow\downarrow}^{\r=1,2}
d\hat\Psi_{\kk,\s,\r}^+
d\hat\Psi_{\kk,\s,\r}^-\Big]\prod_{\kk\in\BBB_{\b,L}^*}
\prod_{\s=\uparrow\downarrow}^{\r=1,2} \hat\Psi^-_{\kk,\s,\r}
\hat\Psi^+_{\kk,\s,\r}=1\label{2.1}\ee
Let us define the free propagator matrix $\hat g_\kk$ as
\be \hat g_\kk=\c_0(2^{-M}|k_0|) \begin{pmatrix}-i k_0 &
-v_0\O^*(\vec k) \cr -v_0\O(\vec k) & -i k_0\end{pmatrix}^{-1}
\label{2.2}\ee
and the ``Gaussian integration'' $P(d\psi)$ as
\bea P(d\Psi) &=& \Big[\prod_{\kk\in\BBB_{\b,L}^*}
^{\s=\uparrow\downarrow}\frac{-\b^2 L^4\,[\c_0(\g^{-M}|k_0|)]^{2}}
{k_0^2+v_0^2|\O(\vec k)|^2} d\hat\Psi_{\kk,\s,1}^+
d\hat\Psi_{\kk,\s,1}^-d\hat\Psi_{\kk,\s,2}^+
d\hat\Psi_{\kk,\s,2}^-\Big]\cdot\nn\\&&\hskip3.truecm \cdot\;\exp
\Big\{-(\b L^2)^{-1}
\sum_{\kk\in\BBB_{\b,L}^*}^{\s=\uparrow\downarrow}
\hat\Psi^{+}_{\kk,\s}\,{\hat
g}_\kk^{-1}\hat\Psi^{-}_{\kk,\s} \Big\}\;. \label{2.3}\eea
Let us also introduce the generating function
\be \WW_{M,\b,L}(A,\phi)=\log\int P(d\Psi)
e^{\VV(\Psi)+(\Psi,\phi)+(A,J)}\label{poras} \ee
where
\bea && \VV(\Psi) =-U\sum_{\r=1,2}\int_{(\b,L)}\!\!\!
d\xx \, \Psi^+_{\xx,\uparrow,\r}
\Psi^-_{\xx,\uparrow,\r}\Psi^+_{\xx,\downarrow,\r}
\Psi^-_{\xx,\downarrow,\r}\;,\label{2.3a}\\
&& (\Psi,\phi)=\sum_{\s=\uparrow\downarrow}
\int_{(\b,L)}\!\!\! d\xx\, \big(\Psi^+_{\xx,\s}\phi_{\xx,\s}^- +
\phi^+_{\xx,\s}\Psi^-_{\xx,\s}\big)\;,\label{2.3b}\\
&&(A,J)=\sum_{\t=\pm}\int_{(\b,L)}\!\!\! d\xx\,
A_{\xx,\t} J^\t_{\xx}+v_0\sum_{j=1,2,3}\int_{(\b,L)}\!\!\! d\xx\,
A_{\xx,j}J^j_\xx\;,\label{2.3c}\eea
and, in the last line, denoting by $\s_1,\s_2,\s_3$ the standard Pauli
matrices,
\be \s_1=\Biggl(\begin{matrix}0&1\\1&0\end{matrix}\Biggr)\;,\qquad
\s_2=\Biggl(\begin{matrix}0&-i\\i&0\end{matrix}\Biggr)\;,\qquad
\s_3=\Biggl(\begin{matrix}1&0\\0&-1\end{matrix}\Biggr)\label{sigma.mat}\ee
and defining $n_{\pm}=(1\pm\s_3)/2$ and $\s_\pm=(\s_1\pm i\s_2)/2$,
\be J^\pm_{\xx}=\sum_{\s=\uparrow\downarrow}\Psi^+_{\xx,\s}n_\pm
\Psi^-_{\xx,\s}\;,\qquad
J^j_\xx=\frac{2i}3\sum_{\s=\uparrow\downarrow}
\Big[\Psi^+_{\xx,\s}\s_+\Psi^-_{\xx+\dd_j,\s}-
\Psi^+_{\xx+\dd_j,\s}\s_-\Psi^+_{\xx,\s}\Big]\;.\label{2.3de}\ee
We will be particularly concerned with the three-points functions:
\be \lis G^{M,\b,L}_{2,1;\sharp}(\zz;\xx,\yy)=e\frac{\dpr}{\dpr A_{\zz,\sharp}}
\frac{\dpr^2}{\dpr
\phi^-_{\yy,\s}\dpr\phi^+_{\xx,\s}} \WW_{M,\b,L}(A,\phi)\Big|_{A=\phi=0}\;,
\qquad \sharp\in\{+,-,1,2,3\}\;,
\label{2.3e}\ee
and with the current-current response functions:
\be {\overline{K}}^{M,\b,L}_{\sharp,\,\flat}(\xx;\yy)=e^2
\frac{\dpr^2}{\dpr A_{\xx,\sharp}\dpr A_{\yy,\flat}}
\WW_{M,\b,L}(A,\phi)\Big|_{A=\phi=0}\;,\qquad \sharp,\flat\in\{+,-,1,2,3\}\;.
\label{2.3f}\ee
The connection between these functions and the corresponding objects evaluated
in he Hamiltonian model of Section \ref{sec1a} is
provided by the following proposition, which is the analogue of Proposition 1
of \cite{GM}.

\vskip.3cm
{\bf Proposition 2.} {\it For any $\b,L<+\io$,
assume that there exists $U_0$ independent of $\b$ and $L$ such that
the three-points functions $G^{M,\b,L}_{2,1;\sharp}(\zz;\xx,\yy)$
and current-current response functions $K^{M,\b,L}_{\sharp,\,\flat}(\xx;\yy)$
at distinct
space-time points are analytic in the complex domain $|U|\le U_0$,
uniformly convergent as $M\to\io$. Then, if $|U|\le U_0$ and
$\zz,\xx,\yy$ are three distinct space-time points,
\bea
e\media{\r^\t_{\zz};\psi^-_{\xx,\s}\psi^+_{\yy,\s}}_{\b,L}&=&
\lim_{M\to\io}\lis G^{M,\b,L}_{2,1;\t}(\zz;\xx,\yy)\;,\qquad \t=\pm\;,\\
ev_0\media{J^j_\zz;\psi^-_{\xx,\s}\psi^+_{\yy,\s}}_{\b,L}&=&
\lim_{M\to\io}\lis G^{M,\b,L}_{2,1;j}(\zz;\xx,\yy)\;,\qquad j=1,2,3\;,\label{3.g21j}\eea
where: the averages in the l.h.s are defined as in Eq.(\ref{1.ave}) and
following lines;
$\r^A_\xx=\sum_\s a^+_{\xx,\s}a^-_{\xx,\s}$ and $\r^B_\xx=\sum_\s
b^+_{\xx+\dd_1,\s}
b^-_{\xx+\dd_1,\s}$. Moreover, if $\xx,\yy$ are two distinct space-time points,
\bea
e^2\media{\r^\t_{\xx};\r^{\t'}_{\yy}}_{\b,L}&=&
\lim_{M\to\io}\lis K^{M,\b,L}_{\t,\t'}(\xx;\yy)\;,\qquad \t,\t'=\pm\;,\\
e^2v_0\media{\r^\t_{\xx};J^j_{\yy}}_{\b,L}&=&
\lim_{M\to\io}\lis K^{M,\b,L}_{\t,j}(\xx;\yy)\;,\qquad \t=\pm\;,\quad
j=1,2,3\;,\\
e^2v_0^2\media{J^j_\xx;J^{j'}_{\yy}}_{\b,L}&=&
\lim_{M\to\io}\lis K^{M,\b,L}_{j,j'}(\xx;\yy)\;,\qquad j,j'=1,2,3\;,\eea
}
The proof of this statement is completely analogous to the one in Appendix B of
\cite{GM} and will not be repeated here. Note that, once that the various
correlation functions in Proposition 2 are known, we can reconstruct the
functions $\hat K^{\b,L}_{\m\n}(\pp)$ and $\hat G^{\b,L}_{2,1;\m}(\kk,\pp)$ in
Eqs.(\ref{1.jj})-(\ref{1.3p}) simply by Fourier transformation (provided
that the correlation functions in Proposition 2 have good enough decay
properties so that the Fourier transform is well defined), e.g.,
\bea && \hat G^{\b,L}_{2,1;0}(\kk,\pp)=\frac{e}{\b L^2}
\int_{(\b,L)}\!\!\!\!\! d\xx\int_{(\b,L)}\!\!\!\!\! d\yy
\int_{(\b,L)}\!\!\!\!\! d\zz\,e^{i\kk(\xx-\yy)+i\pp(\xx-\zz)}
\media{(\r^A_{\zz}+e^{-i\vec p\vec\d_1}\r^B_{\zz+\dd_1});\psi^-_{\xx,\s}
\psi^+_{\yy,\s}}_{\b,L}\;,\nn\\
&& \hat G^{\b,L}_{2,1;l}(\kk,\pp)=\frac{ev_0}{\b L^2}
\int_{(\b,L)}\!\!\!\!\! d\xx\int_{(\b,L)}\!\!\!\!\! d\yy
\int_{(\b,L)}\!\!\!\!\! d\zz\,e^{i\kk(\xx-\yy)+i\pp(\xx-\zz)}
\sum_{j=1}^3\media{(\vec\d_j)_l\h^j_{\vec p}J^j_\zz;\psi^-_{\xx,\s}
\psi^+_{\yy,\s}}_{\b,L}\;;\label{3.g21}\eea
similar formulas are valid for $\hat K^{\b,L}_{\m\n}(\pp)$.

\subsection{Renormalization Group}

The naive perturbative expansion in $U$ of $\WW_{M,\b,L}(A,\phi)$ is
affected by infrared divergences due to the singularity at the Fermi points
of the free propagator $S_0(\kk)$, see Eq.(\ref{11}). The case with $A=0$ has
been studied in detail in \cite{GM}, where it has been shown that the apparent
divergences affecting the naive perturbation series can be cured by proper
iterative resummations; these allowed us to recast the original expansion into
a new convergent expansion (uniformly in $M,\b,L$) involving a sequence of
effective parameters $Z_h,v_h$, playing the role of effective wave function
renormalization and Fermi velocity at momentum scale $2^h$, $h\le 0$,
relative to the Fermi points. The iterative resummations can be implemented
by using constructive fermionic Renormalization Group (RG) techniques; a key
point of the analysis, which make the construction of the ground state
possible, is the fact that density-density interactions are {\it irrelevant}
in a RG sense. In this section we review the iterative integration scheme
used to compute $\WW_{M,\b,L}(A,\phi)$, with particular emphasis to the
novelties due to presence of the external field $A$.

\subsubsection{The ultraviolet integration.}\label{b1}

Proceeding as in Sections 3.2 and 3.4 of
\cite{GM}, we decompose the
propagators $\hat{g}(\kk)$ into sums of two propagators supported
in the regions of $k_0$ ``large'' and ``small''. The regions of
$k_{0}$ large and small are defined in terms of the smooth support
function $\chi(t)$; the constant $a_0$ entering in its definition
is chosen so that the supports of $\chi_0\Big( \sqrt{k_0^2 + |\vec k
- \vec p_{F}^{+}|} \Big)$ and $\chi_0\Big( \sqrt{k_0^2 + |\vec k -
\vec p_{F}^{-}|} \Big)$ are disjoint (e.g., $a_0=1/3$ is fine).
We rewrite $\hat{g}(\kk)$ as
\be \hat{g}(\kk) = \hat{g}^{(u.v.)}(\kk) + \hat
g^{(i.r.)}(\kk)\;\label{1.2.37} \ee
where, setting $\pp_{F}^{\o} = (0,\vec p_{F}^{\o})$ with $\o=\pm$:
\be \hat{g}^{(u.v.)}(\kk) = \hat{g}(\kk) - \hat
g^{(i.r.)}(\kk)\;,\qquad g^{(i.r.)}(\kk) = \sum_{\o =
\pm}\chi_0(|\kk - \pp_{F}^{\o}|)\hat{g}(\kk)\;,\label{1.2.38} \ee
Defining $V(A,\Psi,\phi)=\VV(\Psi)+(\Psi,\phi)+(A,J)$, we can rewrite
%
\be e^{\WW_{M,\b,L}(A,\phi)}=\int P(d\Psi) e^{V(A,\Psi,\phi)}=
\int P(d\Psi^{(i.r.)})\int
P(d\Psi^{(u.v.)})\eu^{V(A,\Psi^{(i.r.)} + \Psi^{(u.v.)},\phi)}\;,
\label{2.3.33}\ee
where $P(d\Psi^{(i.r.)})$ and $P(d\Psi^{(u.v.)})$ are the Gaussian integrations
associated to the propagators $\hat{g}^{(i.r.)}(\kk)$ and
$\hat{g}^{(u.v.)}(\kk)$, respectively.
Using Eq.(\ref{2.3.33}) we can further rewrite the generating functional as:
\bea  e^{\WW_{M,\b,L}(A,\phi)}&=&
\int P(d\Psi^{(i.r.)})\exp\,\big\{\, \sum_{n\ge
1}\frac{1}{n!}\EE^T_{u.v.}(V(A,\Psi^{(i.r.)}+\cdot,\phi);n)\big\}=
\label{2.2.14}\\
&=:&
e^{-\b L^2 F_{0,M}}
\int P(d\Psi^{(i.r.)}) e^{\VV_M(\Psi^{(i.r.)})
+B_M(A,\Psi^{(i.r.)},\phi)+(\phi,\Psi^{(i.r.)})+(A,J^{(i.r.)})}\;,
\nn\eea
where: $\EE^T_{u.v.}$ is the truncated expectation with respect to the
propagator
$\hat g^{(u.v.)}(\kk)$; $F_{0,M}$ is a constant;
$\VV_M$ is the effective potential on scale $0$;
$B_M(A,\Psi^{(i.r.)},\phi)$ collects the terms depending
on $A,\phi$ generated by the ultraviolet integration; $J^{(i.r.)}$ is defined
in the same way as $J$
(see Eq.(\ref{2.3de})) with $\Psi$ replaced by $\Psi^{(i.r.)}$. As proved in
\cite{GM} (see Eq.(3.36)
and Lemma 2 of \cite{GM}), the effective potential $\VV_M$ can be written as
\bea && {\cal V}_M(\Psi^{(i.r.)})=\sum_{n=1}^\io (\b L^2)^{-2n}
\sum_{\s_1,\ldots,\s_n=\uparrow
\downarrow}\sum_{\r_1,\ldots,\r_{2n}=1,2}\sum_{\kk_1,\ldots,\kk_{2n}}
\Big[\prod_{j=1}^n\hat \Psi^{(i.r.)+}_{\kk_{2j-1},\s_j,\r_{2j-1}}
\hat \Psi^{(i.r.)-}_{\kk_{2j},\s_j,\r_{2j}}\Big]\;\cdot\nn\\
&&\hskip5.truecm\cdot\hat W_{M;2n;\ul\r}(\kk_1,\ldots,
\kk_{2n-1})\;\d(\sum_{j=1}^n(\kk_{2j-1}-\kk_{2j}))\;,\label{2.2.16}\eea
where $\ul\r=(\r_1,\ldots,\r_{2n})$ and we used the notation
\be \d(\kk)=\d(\vec k)\d(k_0)\;,\qquad \d(\vec k)=L^2\sum_{n_1,n_2\in\zzz}
\d_{\vec k,n_1\vec G_1+n_2\vec G_2}\;,\qquad \d(k_0)=\b\d_{k_0,0}\;,
\label{2.16a}\ee
with $\vec G_1,\vec G_2$ the basis of $\L^*$ defined after Eq.(\ref{1.f}).
Moreover
(see Lemma 2 of \cite{GM}),
the constant $F_{0,M}$ and the kernels $\hat W_{M;2n;\ul\r}$ are given by
power series in $U$, convergent in the complex disc $|U|\le U_0$, for $U_0$
small enough and independent of $\b,L,M$; after Fourier
transform, the $\xx$-space counterparts of the kernels $\hat W_{M;2n;\ul\r}$
satisfy the following bounds:
\be \int d\xx_1\cdots d\xx_{2n}\Big[\prod_{1\le i<j\le 2n}
||\xx_i-\xx_j||^{m_{i,j}}\Big]\big|W_{M;2n;\ul\r}(\xx_1,\ldots,\xx_{2n})\big|
\le \b|\L|C_{m}^n |U|^{\max\{1,n-1\}}\;,\label{2.2.17}\ee
for some constant $C_m>0$, where $m=\sum_{1\le i<j\le 2n}m_{i,j}$.
The limits $F_0=\lim_{M\to\io}F_{0,M}$ and
$W_{2n;\ul\r}
(\xx_1,\ldots,\xx_{2n})=\lim_{M\to\io}W_{M;2n;\ul\r}
(\xx_1,\ldots,\xx_{2n})$ exist and are reached uniformly in $M$, so that,
in particular, the limiting functions are analytic in the same domain $|U|\le
U_0$.

The functional $B_M(A,\Psi^{(i.r.)},\phi)$ admits very similar representation
and bounds, i.e.,
\bea  B_M(A,\Psi^{(i.r.)},\phi)&=&\sum_{n,m=0}^\io \frac1{(\b L^2)^{2n+m}}
\sum_{\substack{\ul\s,\ul\r\\\ul\sharp,\ul\g}}\sum_{\ul\kk,\ul\pp}
\Big[\prod_{j=1}^n\hat \f^{+}_{\kk_{2j-1},\s_j,\r_{2j-1},\g_{2j-1}}
\hat \f^{-}_{\kk_{2j},\s_j,\r_{2j},\g_{2j}}\Big] \Big[ \prod_{i=1}^{m}
\hat A_{\sharp_{i},\pp_{i}}\Big]\nn\\
&&\cdot\hat W_{M;2n,m;\ul\r,\ul\sharp;\ul\g}(\{\kk_{j}\},\{\pp_{i}\})\,\d\Big(
\sum_{j=1}^{n}\big(\kk_{2j-1} - \kk_{2j}\big) -
\sum_{i=1}^{m}\pp_{i}\Big)\;, \label{2.2.16a}\eea
where $\underline{\s} :=
(\s_{1},\ldots,\,\s_{n})$, $\underline{\sharp} :=
(\sharp_{1},\ldots,\,\sharp_{m})$ (with $\sharp_i\in\{+,-,1,2,3\}$),
$\ul\g=(\g_1,\ldots,\g_{2n})$ (with $\g_i\in\{ext,int\}$),
$\ul\kk=(\kk_1,\ldots,\kk_{2n})$, $\ul\pp=(\pp_1,\ldots,\pp_m)$
and the field $\hat\f$ can be either $\Psi^{(i.r.)}$ or $\phi$, depending on
the label
$\g$, i.e., $\hat\f^\pm_{\kk,\s,ext}=\hat\phi^\pm_{\kk,\s}$ and
$\hat\f^\pm_{\kk,\s,int}=\hat\Psi^{(i.r.)\pm}_{\kk,\s}$. The kernels
$\hat W_{M;2n,m;\ul\r,\ul\sharp;\ul\g}$ are analytic in $U$, they
admit bounds analogous to
Eq.(\ref{2.2.17}), uniformly in $\ul\g$ and $M$, and converge uniformly as
$M\to\io$ to
limiting kernels denoted by $\hat W_{2n,m;\ul\r,\ul\sharp;\ul\g}$ (for this
reason, the label $M$ will not play any important role in the following and
will be dropped from
now on). The proof of these claim goes along the same lines as the proof of
Lemma 2 in \cite{GM}
and will not repeated here.

The kernels $\hat W_{2n,m;\ul\r,\ul\sharp;\ul\g}$ satisfy a number of symmetry
properties inherited from
the symmetries of the action and of the Gaussian integration, described and
proved in Appendix
\ref{secB}. In particular, thinking of $\hat W_{2n,m;\ul\sharp;\ul\g}$ as
tensors with entries $\hat W_{2n,m;\ul\r,\ul\sharp;\ul\g}$
and defining $\hat W_{2,\o}(\kk'):=W_{2,0;(int,int)}(\kk' + \pp_{F}^{\o},
\kk'+\pp_{F}^{\o})$, $\hat W_{\sharp,\flat}(\pp):=\hat W_{0,2;\sharp,\flat}
(\pp,-\pp)$, $\hat W_{\sharp,\o}(\kk',\pp):=\hat W_{2,1;\sharp;(int,int)}
(\kk'+\pp_F^\o+\pp,\kk'+\pp_F^\o,\pp)$,
\bea\hat W_2(\kk')&=&-iz_0k_0'+z_1(k_1'\s_2+
\o k_2'\s_1)+ O(|\kk'|^{2})\;,\label{1.2.49}\\
\hat W_{\t,\t'}(\pp)&=&e^{-i\frac{\vec p}2\vec\d_1(\t-\t')}(a+a'\t\t')
+O(|\pp|^2)\;,\label{1.2.50a}\\
\hat W_{\t,j}(\pp)&=&b\t p_0+O(|\pp|^2)\;,\label{1.2.50b}\\
\hat W_{j,j'}(\pp)&=&e^{i\frac{\vec p}{2}(\vec\d_j-\vec\d_{j'})}
(c\d_{j,j'}+c')+O(|\pp|^2)\;,\label{1.2.50c}\\
\hat W_{\t,\o}(\kk',\pp)&=&
\l_0+\l_3\t\s_3+O(|\kk'|+|\pp|)\;,\label{1.2.51a}\\
\hat W_{j,\o}(\kk',\pp)&=&\l_1 e^{i\o\frac{2\p}{3}(j-1)\s_3}\s_2+
O(|\kk'|+|\pp|)\;,\label{1.2.51b}\eea
where the constants $z_0,z_1,a,a',b,c,c',\l_0,\l_1,\l_3$ are all real.
The proof of Eq.(\ref{1.2.49}) is in Lemma 3 of \cite{GM}, while the
proof of the remaining relations is in Appendix \ref{secD2}.

\subsubsection{The infrared integration.}\label{b2}

After the integration of the ultraviolet modes, we decompose the infrared
propagator as a sum of two quasi-particle propagators:
\be g^{(i.r.)}(\xx-\yy)=\sum_{\o=\pm} e^{-i \pp_F^{\;\o}(\xx-\yy)}
g_{\o}^{(\le 0)}(\xx-\yy)\;,\label{2.22}\ee
where, if $\kk'=(k_0,\vec k')$,
\be g_{\o}^{(\le 0)}(\xx-\yy)=\frac1{\b L^2}\sum_{\kk'\in\BBB^\o_{\b,L}}
\chi_0(|\kk'|)e^{-i\kk'(\xx-\yy)}\begin{pmatrix}-i k_0 & -v_0\O^*(\vec k'+
\vec p_F^{\;\o})
\\ -v_0\O(\vec k'+\vec p_F^{\;\o}) & -i k_0\end{pmatrix}^{-1}\label{2.23}\ee
and $\BBB_{\b,L}^\o=\big\{2\p\b^{-1}(\ZZZ+\frac12)\cap
\{k_0\,:\,\c_0(2^{-M}|k_0|)>0\}\big\}
\times\{\frac{n_1}L\vec G_1
+\frac{n_2}L\vec G_2-\vec p_F^{\;\o}\;,\ 0\le n_1,n_2<L\}$.
Correspondingly, we rewrite $\Psi^{(i.r.)}$ and $\phi^\pm$
as sums of two independent Grassmann fields, $\Psi^{(i.r.)\pm}_{\xx,\s}=
\sum_{\o=\pm}e^{\pm i\pp_F^\o\xx}\Psi^{(\le 0)\pm}_{\xx,\s,\o}$ and
$\phi^{\pm}_{\xx,\s}=\sum_{\o=\pm}e^{\pm i\pp_F^\o\xx}\phi^{\pm}_{\xx,\s,\o}$,
and rewrite the generating functional as (dropping systematically the
$M$ label):
\bea &&e^{\WW_{\b,L}(A,\phi)}= \label{3.1aa}\\
&&\quad =e^{-\b L^2 F_0+S^{(\ge 0)}(A,\phi)}
\int P_{\c_0,C_0}(d\Psi^{(\le 0)})
e^{{\cal V}^{(0)}(\Psi^{(\le 0)})+B^{(0)}(A,\Psi^{(\le 0)},\,\phi)+
 (A \Psi^{(\le 0)},{T_1}\Psi^{(\le 0)})+(\phi, \Psi^{(\le 0)})}\;,\nn\eea
where:
\bea &&P_{\c_0,C_0}(d\Psi^{(\le 0)})={{\cal N}_0}^{-1}\Biggl[\;
\prod_{\kk'\in\BBB_{\b,L}^\o}^{\c_0(|\kk'|)>0}
\;\prod_{\s,\o,\r}
d\hat\Psi^{(\le 0)+}_{\kk',\s,\r,\o}d\hat\Psi^{(\le 0)-}_{\kk',\s,\r,\o}\Biggr]
\cdot\label{2.26}\\
&&\hskip2.truecm\cdot
\exp \Big\{-(\b L^2)^{-1}\sum_{\o=\pm}^{\s=\uparrow\downarrow}
\sum_{\kk'\in\BBB_{\b,L}^\o}^{\c_0(|\kk'|)>0}\c_0^{-1}(|\kk'|)
\hat\Psi^{(\le 0)+}_{\kk',\s,\o}C_{0,\o}(\kk')\hat\Psi^{(\le 0)-}_{
\kk',\s,\o}\}\nn\eea
and
\bea
C_{0,\o}(\kk')&=&\begin{pmatrix}-i k_0 & -v_0\O^*(\vec k'+\vec p_F^{\;\o})
\\ -v_0\O(\vec k'+\vec p_F^{\;\o}) & -i k_0\end{pmatrix}=\label{2.a0}\\
&=:&Z_0
\begin{pmatrix}-i k_0  & v_0(ik_1'-\o k_2')
\\ v_0(-ik_1'-\o k_2') & -i k_0 \end{pmatrix}(1+R_{0,\o}(\kk'))
\;,\nn\eea
with ${\cal N}_0$ chosen in such a way that $\int P_{\c_0,C_0}
(d\Psi^{(\le 0)})=1$, $Z_0=1$ and $R_{0,\o}$ a matrix such that
$||R_{0,\o}(\kk')||\le C|\kk'|^2$. Moreover, the functionals
$\VV^{(0)}$, $(A \Psi^{(\le 0)},{T_1}\Psi^{(\le 0)})$ and $(\phi,\Psi^{(\le 0)})$ are the
same as $\VV_M$, $(A,J^{(i.r.)})$ and $(\phi,\Psi^{(i.r.)})$ in
Eq.(\ref{2.2.14}), with $\Psi^{(i.r.)}$ and $\phi^\pm$ rewritten in terms of
$\Psi^{(\le 0)}_\o$ and $\phi^\pm_\o$; similarly $S^{(\le 0)}(A,\phi)+
B^{(0)}$ is the same as $B_M$ after the same rewriting, with $S^{(\le 0)}$
collecting all the terms depending on $A$ and/or $\phi$ but not on
$\Psi^{(\le 0)}$. E.g., to be more explicit, defining
$\hat A^{\o-\o'}_{\pp',\sharp}:=\hat A_{\pp_F^\o-\pp_F^{\o'}+\pp',\sharp}$
and recalling the definition of $n_\pm$, i.e., $n_\pm=(1\pm\s_3)/2$,
%
\be (A \Psi^{(\le 0)},{T_1}\Psi^{(\le 0)})=
\frac1{(\b L^2)^2}\sum_{\o,\o',\s,\sharp}\sum_{\kk',\pp'}
\hat A^{\o-\o'}_{\pp',\sharp}\hat\Psi^{(\le h)+}_{\kk'+\pp',\s,\o}
T_{\sharp,1}^{\,\o,\o'}\!\!(\kk',\pp')
\hat\Psi^{(\le h)-}_{\kk',\s,\o'}\;,\label{3.t0}\ee
with $T_{\t,1}^{\o,\o'}\!\!(\kk',\pp')=n_\t$ and
\be
T_{j,1}^{\o,\o'}\!\!(\kk',\pp')=v_0\,
\frac{2i}{3}\,e^{i\frac{2\p}3(j-1)(\o'n_+-\o n_-)}
\big(\s_+e^{-i\kk'(\dd_j-\dd_1)}-
\s_-e^{+i(\kk'+\pp')(\dd_j-\dd_1)}\big)\;;\label{3.tt00}\ee
similar expressions are valid for the other functionals appearing in the
exponent of Eq.(\ref{3.1aa}).\\

The integration of Eq.(\ref{3.1aa}) is performed in an iterative fashion,
justified by the infrared divergence caused by the singularity at
$\kk'=\V0$ of the propagator $\hat g^{(\le 0)}_\o(\kk')$, see
Eqs.(\ref{2.23}),(\ref{2.a0}). We define
$f_h(\kk'):=\c_0(2^{-h}|\kk'|)
-\c_0(2^{-h+1}|\kk'|)$ and rewrite
\be \c_0(|\kk'|)=\sum_{k=-\io}^0 f_k(\kk')=:\c_h(\kk')+f_{h+1}(\kk')+\cdots+
f_0(\kk')\;.\label{2.27}\ee
The integration procedure consists in the following: we first rewrite
$\c_0=\c_{-1}+f_0$ and correspondingly decompose the propagator
$g^{(\le 0)}_\o=g^{(\le -1)}_\o+g^{(0)}_\o$ and the field $\Psi^{(\le 0)}_\o=
\Psi^{(\le -1)}_\o+\Psi^{(0)}_\o$; next, we integrate out the ``$0$-mode'',
i.e., the Grassmann field $\Psi^{(0)}_\o$, and re-express the result in terms
of a new effective potential on scale $-1$; then we iterate, by integrating out
step by step the degrees of freedom on scale $-1,-2,\ldots,h+1$, with $h<0$.
After the integration of the fields on scales $-1,-2,\ldots,h+1$, we get:
\bea  e^{\WW_{\b,L}(A,\phi)}&=& e^{-\b L^2F_h+S^{(\ge h)}(A,\phi)}
\cdot\label{3.1a}\\&&\cdot\int P_{\c_h,C_h}(d\Psi^{(\le h)})
e^{{\cal V}^{(h)}(\Psi^{(\le h)})+B^{(h)}(A,\Psi^{(\le h)},\phi)+
(A\Psi^{(\le h)},T_{h+1}\Psi^{(\le h)})+(\phi, Q^{(h+1)}\Psi^{(\le h)})}\;,\nn\eea
where $F_h$, $C_h$, ${\cal V}^{(h)}$, $B^{(h)}$, $T_h$, $Q^{(h)}$
will be defined recursively (in particular, $F_h$, $C_h$ [called $A_h$ in \cite{GM}],
${\cal V}^{(h)}$,
$Q^{(h)}$ have already been defined in \cite{GM})
and $P_{\c_h,C_h}(d\Psi^{(\le
h)})$ is defined in the same way as $P_{\c_0,C_0}(d\Psi^{(\le 0)})$ with
$\Psi^{(\le 0)}, \c_0, C_{0,\o}, Z_0, v_0, R_{0,\o}$ replaced
by $\Psi^{(\le h)}, \c_h, C_{h,\o}, Z_h, v_h, R_{h,\o}$,
respectively.
Moreover,
\bea  {\cal V}^{(h)}(\Psi)&=&\sum_{n=1}^\io (\b L^2)^{-2n}
\sum_{\ul\s,\ul\r,\ul\o,\ul\kk'}\;
\Big[\prod_{j=1}^n\hat\Psi^{(\le h)+}_{\kk_{2j-1}',\s_j,\r_{2j-1},\o_{2j-1}}
\hat \Psi^{(\le h)-}_{\kk_{2j}',\s_j,\r_{2j},\o_{2j}}\Big]\cdot\nn\\
&&\cdot\hat W_{2n;\ul\r;\ul\o}^{(h)}(\kk_1',\ldots,
\kk_{2n-1}')\;\d(\sum_{j=1}^{2n}(-1)^j(\pp_F^{\o_{j}}+\kk_{j}'))
\label{2.29}\eea
and
\bea  &&
B^{(h)}(A,\Psi,\phi)=\sum_{n,m\ge 0}^* (\b L^2)^{-2n-m}\!\!\!
\sum_{\substack{\ul\s,\ul\r,\ul\sharp,\ul\g,\ul\o\\ \ul\kk',\ul\pp'}}^*
\Big[\prod_{j=1}^n\hat \f^{+}_{\kk_{2j-1}',\s_j,\r_{2j-1},\g_{2j-1},\o_{2j-1}}
\hat \f^{-}_{\kk_{2j}',\s_j,\r_{2j},\g_{2j},\o_{2j}}\Big]\hskip1.truecm{\phantom{.}}
\label{2.2.16abc}\\ &&
\cdot\Big[ \prod_{i=1}^{m}\hat A_{\sharp_{i},\pp_{i}'}^{\o_i-\o_i'}\Big]
\hat W^{(h)}_{2n,m;\ul\r,\ul\sharp;\ul\o;\ul\g}(\{\kk_{j}'\},\{\pp_{i}'\})\,
\d\Big(\sum_{j=1}^{2n}(-1)^{j+1}\big(\kk_j'+\pp_F^{\o_j}\big) -\sum_{i=1}^m
(\pp_{i}'+\pp_F^{\o_i}-\pp_F^{\o_i'})\Big)\nn\eea
where: the $*$'s on the sums remind the fact that only terms explicitly
depending on $\hat\f_{int}^\pm=\hat \Psi^{(\le h)\pm}$ contribute to
$B^{(h)}$; $\ul\o=(\o_1,\ldots,\o_{2n};\o_1-\o_1',\ldots,\o_m-\o_m')$.

The terms $(A \Psi^{(\le h)},{T_{h+1}}\Psi^{(\le h)})$ and $(\phi, Q^{(h+1)}\Psi^{(\le h)})$ read:
\bea && 
(A \Psi^{(\le h)},{T_{h+1}}\Psi^{(\le h)})=
\frac1{(\b L^2)^2}\sum_{\substack{\o,\o',\s,\sharp\\
\kk',\pp'}}
\hat A^{\o-\o'}_{\pp',\sharp}\hat\Psi^{(\le h)+}_{\kk'+\pp',\s,\o}
T_{\sharp,h+1}^{\,\o,\o'}\!(\kk',\pp')
\hat\Psi^{(\le h)-}_{\kk',\s,\o'}\;,\label{2.2.aj}\\
%
&& (\phi, Q^{(h+1)}\Psi^{(\le h)})=
\frac1{\b L^2}\sum_{\o,\s,\kk'}
\Big(\hat \Psi^{(\le h)+}_{\kk',\s,\o}\hat Q^{(h+1)-}_{\kk',\o}
\hat \phi^{-}_{\kk',\s,\o}+\hat \phi^{+}_{\kk',\s,\o}\hat Q^{(h+1)+}_{\kk',\o}
\hat \Psi^{(\le h)-}_{\kk',\s,\o}\Big)\;,\label{2.2.q}\eea
for suitable matrices $T^{\o,\o'}_{\sharp,h+1}(\kk',\pp')$ and $\hat
Q^{(h+1)\pm}_{\kk',\o}$, with $Q^{(1)\pm}_{\kk',\o}=1$.
The iterative procedure goes on up to scale $h_\b$,
where $h_\b$ is the largest scale such that $a_0 \g^{h_\b-1}<\frac{\p}{\b}$. 
The result of the last iteration is $e^{\WW_{\b,L}(A,\phi)}$.

\subsubsection{Localization and renormalization.}\label{b3}

In order to inductively prove Eq.(\ref{3.1a}), we rewrite
\be {\cal V}^{(h)}(\Psi^{(\le h)}) =\LL{\cal V}^{(h)}(\Psi^{(\le h)})+\RR{\cal V}^{(h)}(\Psi^{(\le h)})
\;,\label{2.loc}\ee
where
\be
\LL{\cal V}^{(h)}(\Psi^{(\le h)})=
\frac1{\b L^2}\sum_{\s,\o}\;
\sum_{\kk'} \hat\Psi^{(\le h)+}_{\kk',\s,\o}
\hat W_{2;(\o,\o)}^{(h)}(\kk')
\hat\Psi^{(\le h)-}_{\kk',\s,\o}\;,\label{2.30a}\ee
and $\RR{\cal V}^{(h)}$ is given by \pref{2.29} with
$\sum_{n=1}^\io$ replaced by $\sum_{n=2}^\io$, that is it contains
only the monomials with four or more fields (note that in Eq.(\ref{2.30a})
the two fermionic fields have the same $\o$-index; terms with two different
quasi-particles indices are not allowed by momentum conservation, see also
the remark after (3.62) of \cite{GM}). Moreover, defining
\bea&&
\lis W^{(h)-}_{2,\o}(\kk'):=\hat W^{(h)}_{2,0;(\o,\o);(int,ext)}(\kk',\kk')\;,
\qquad
\lis W^{(h)+}_{2,\o}(\kk'):=\hat W^{(h)}_{2,0;(\o,\o);(ext,int)}(\kk',\kk')
\;,\nn\\
&&\hat W^{(h)}_{\sharp;\o,\o'}(\kk',\pp'):=\sum_{\o_1,\o_2}
\hat W_{2,1;\sharp;(\o,\o';\o-\o');(int,int)}^{(h)}(\kk'+\pp',\kk',\pp')\;,\label{2.ciao}\eea
we rewrite:
\be B^{(h)}(A,\Psi^{(\le h)},\phi)=\LL B^{(h)}(A,\Psi^{(\le h)},\phi)+\RR B^{(h)}(A,\Psi^{(\le h)},\phi)
\;,\label{2.locB}\ee
where
\bea \LL B^{(h)}(A,\Psi^{(\le h)},\phi)&=&
\frac1{\b L^2}\sum_{\s,\o,\kk'}
\Big(\hat \Psi^{(\le h)+}_{\kk',\s,\o}\lis W^{(h)-}_{2,\o}
(\kk')\hat\phi^-_{\kk',\s,\o}+
\hat \phi^{+}_{\kk',\s,\o}\lis W^{(h)+}_{2,\o}
\hat\Psi^-_{\kk',\s,\o}\Big)+\nn\\
&+&\frac1{(\b L^2)^2}\sum_{\substack{\o,\o'\\ \s,\sharp}}
\,\sum_{\kk',\pp'}
\hat A^{\o-\o'}_{\pp',\sharp}
\hat\Psi^{(\le h)+}_{\kk'+\pp',\s,\o}
\hat W_{\sharp;\o,\o'}^{(h)}(\kk',\pp')
\hat\Psi^{(\le h)-}_{\kk',\s,\o}\;.\label{2.30aab}\eea
At this point we ``reabsorbe" $\LL\VV^{(h)}$ in the fermionic gaussian integration
and $\LL B^{(h)}$ into the definition of the effective source terms:
\bea &&e^{-\b L^2 F_h+S^{(\ge h)}(A,\phi)}\int P_{\c_h,C_h}(d\Psi^{(\le h)})
e^{{\cal V}^{(h)}(\Psi^{(\le h)})+B^{(h)}(A,\Psi^{(\le h)},\phi)+
(A\Psi^{(\le h)},T_{h+1}\Psi^{(\le h)})+(\phi, Q^{(h+1)}\Psi^{(\le h)})}\nn\\
&&\qquad\qquad=e^{-\b L^2(F_h+ e_h)+S^{(\ge h)}(A,\phi)}\int  P_{\c_h,\lis C_{h-1}}(d\Psi^{(\le h)})e^{\RR\VV^{(h)}(\Psi^{(\le h)})}\cdot\label{2.31aba}\\
&&\hskip6.truecm\cdot e^{\RR B^{(h)}(A,\Psi^{(\le h)},\phi)
+(A \Psi^{(\le h)},{T_{h}}\Psi^{(\le h)})
+(\phi, Q^{(h)}\Psi^{(\le h)})}\;,\nn\eea
where $e_h$ is a suitable constant (see Eq.(3.67) of \cite{GM}) and
\bea && \lis C_{h-1,\o}(\kk')=C_{h,\o}(\kk')+\c_h(\kk')\hat W_{2;(\o,\o)}^{(h)}(\kk')\;,\nn\\
&& \hat Q^{(h)-}_{\kk',\o}=\hat Q^{(h+1)-}_{\kk',\o}+\hat W^{(h)}_{2;(\o,\o)}(\kk')\sum_{k=h+1}^1
\hat g^{(k)}_{\o}\hat Q^{(k)-}_{\kk',\o}\;,\label{2.31azd}\\
&&\hat Q^{(h)+}_{\kk',\o}=\hat Q^{(h+1)+}_{\kk',\o}+\Big(\sum_{k=h+1}^1
\hat Q^{(k)+}_{\kk',\o}\hat g^{(k)}_{\o}\Big)\hat W^{(h)}_{2;(\o,\o)}(\kk')\;,\nn\\
&& T_{\sharp,h}^{\,\o,\o'}\!(\kk',\pp')=T_{\sharp,h+1}^{\,\o,\o'}(\kk',\pp')
+\hat W^{(h)}_{\sharp;\o,\o'}(\kk',\pp')\;.\nn\eea
The second and third equations in (\ref{2.31azd}) can be proved as in
\cite{GM}, see Eqs.(3.111)-(3.113).
We are now ready to perform the integration of the $\Psi^{(h)}$ field:
we rewrite the Grassmann field $\Psi^{(\le h)}$ as a sum of two independent
Grassmann fields $\Psi^{(\le h-1)}+\Psi^{(h)}$ and correspondingly we rewrite
the r.h.s. of Eq.(\ref{2.31aba}) as
\bea && e^{-\b L^2(F_h+e_h)+S^{(\ge h)}(A,\phi)}
\!\!\!\int P_{\c_{h-1},C_{h-1}}(d\Psi^{(\le h-1)}) \,
\!\!\!\int P_{f_h,\lis C_{h-1}}
(d\Psi^{(h)})e^{\RR{\cal V}^{(h)}(\Psi^{(\le h-1)}+\Psi^{(h)})}\cdot\nn\\
&&\cdot e^{
\RR B^{(h)}(A,\Psi^{(\le h-1)}+\Psi^{(h)},\phi)
+(A(\Psi^{(\le h-1)}+\Psi^{(h)}),T_{h}(\Psi^{(\le h-1)}+\Psi^{(h)}))
+(\phi, Q^{(h)}(\Psi^{(\le h-1)}+\Psi^{(h)}))}\;,\label{2.35}\eea
where
\be C_{h-1,\o}(\kk')=C_{h,\o}(\kk')+\hat W_{2;(\o,\o)}^{(h)}(\kk')\;.\label{CC.3}\ee
In \cite{GM} we proved (see Eqs.(3.68)-(3.69) and proof of Theorem 2 of \cite{GM})
that the single scale propagator, defined as
\bea && \int P_{f_h,\lis C_{h-1}}(d \Psi^{(h)})
\Psi^{(h)-}_{\xx_1,\s_1,\o_1}
\Psi^{(h)+}_{ \xx_2,\s_2,\o_2} = \d_{\s_1,\s_2}\d_{\o_1,\o_2}
g^{(h)}_\o(\xx_1-\xx_2)\;,\nn\\
&& g^{(h)}_\o(\xx_1-\xx_2)=\frac1{\b L^2}\sum_{\kk'\in\BBB_{\b,L}^\o}
e^{-i\kk'(\xx_1-\xx_2)}f_h(\kk')\Big[\lis A_{h-1,\o}(\kk')\Big]^{-1}
\;,\label{2.37}\eea
can be rewritten as
\bea && g^{(h)}_\o(\xx_1-\xx_2)=\frac1{\b L^2}\sum_{\kk'\in\BBB_{\b,L}^\o}
e^{-i\kk'(\xx_1-\xx_2)}\cdot\label{2.37erc}\\
&&\qquad\cdot\frac{f_h(\kk')}{\lis Z_{h-1}(\kk')}\begin{pmatrix}-i k_0  & \lis v_{h-1}(\kk')
(ik_1'-\o k_2')
\\ \lis v_{h-1}(\kk')(-ik_1'-\o k_2')& -i k_0 \end{pmatrix}^{\!\!-1}\!\!(1+\lis R_{h-1,\o}(\kk'))^{-1}
\;,\nn\eea
with $\lis Z_{h-1},\lis v_{h-1}$ two functions such that (choosing $0<\th<1$)
\bea && |\lis Z_{h-1}(\kk')-\lis Z_h(\kk')|\le (\const.)|U| 2^{\th h}\;,\qquad \lis Z_0(\kk')=Z_0=1\;,\nn\\
&& \big|\lis v_{h-1}(\kk')-\lis v_{h}(\kk')|\big|\le (\const.)|U| 2^{\th h}\;,\qquad
\lis v_{0}(\kk')=v_0=\frac32t\label{3.33333}\eea
and $\lis R_{h-1,\o}$ a matrix such that $||\lis R_{h-1,\o}(\kk')||\le (\const.)|\kk'|^2$
and, if $2^{h-1}\le |\kk'|\le 2^{h+1}$, $||\dpr_{\kk'}^nR_{h-1,\o}(\kk')||\le (\const.)
2^{(2-n)h}$. The constants $Z_h:=\lis Z_h(\V0)$ and $v_h:=\lis v_h(\V0)$ play the role of
effective wave function renormalization and Fermi velocity on scale $h$. Similarly, the
local parts of the matrices $T^{\o,\o'}_{\sharp,h}(\kk',\pp')$ play the role of effective vertices; in particular, the ``relativistic" vertex functions, which represent the dominant contribution
in the infrared to the kernel of the three point function,
are defined as (see the subsection {\it The three-point function} below for more details):
\be Z_{0,h}=\sum_{\t=\pm}T^{\o,\o}_{\t,h}(\V0,\V0)\;,\quad Z_{1,h}=-\s_2\sum_{j=1}^3
(\vec\d_j)_1 T^{\o,\o}_{j,h}(\V0,\V0)\;,\quad
Z_{2,h}=-\o\s_1\sum_{j=1}^3
(\vec\d_j)_2 T^{\o,\o}_{j,h}(\V0,\V0)\;.\label{2.zetamu}\ee
From the symmetries discussed in Appendix \ref{secB} and \ref{secD2} (see also
Eqs.(\ref{1.2.51a})-(\ref{1.2.51b})), $Z_{\m,h}$, $\m=0,1,2$, are all real, independent of $\o$ and
proportional to the identity matrix and, therefore, they can be regarded as constants,
as we will do in the following; moreover, $Z_{1,h}=Z_{2,h}$,
$Z_{0,0}=1$, $Z_{1,0}=Z_{2,0}=v_0$ and $|Z_{\m,h-1}-Z_{\m,h}|\le (\const.)|U|2^{\th h}$,
see Eq.(\ref{8.1}) and following discussion for a proof.

Now, going back to Eq.(\ref{2.35}), if we integrate out the field on scale $h$ and define:
\bea && e^{-\b L^2\lis e_{h-1}+S^{(\ge h-1)}(A,\phi)+{\cal V}^{(h-1)}(\Psi^{(\le h-1)})+
B^{(h-1)}(A,\Psi^{(\le h-1)},\phi)}:= \label{3.ciao3}\\
&&\quad=e^{S^{(\ge h)}(A,\phi)}\!\!\int P_{f_h,\lis C_{h-1}}
(d\Psi^{(h)}) e^{\RR{\cal V}^{(h)}(\Psi^{(\le h-1)}+\Psi^{(h)})+\RR B^{(h)}(A,\Psi^{(\le h-1)}+\Psi^{(h)},\phi)}\cdot\nn\\
&&\quad\quad\cdot e^{
(A\Psi^{(h)},T_{h}\Psi^{(\le h-1)})+(A\Psi^{(\le h-1)},T_{h}\Psi^{(h)})+(A\Psi^{(h)},T_{h}
\Psi^{(h)})+(\phi, Q^{(h)}\Psi^{(h)})}\;,\nn\eea
we get Eq.(\ref{3.1a}) with $h$ replaced by $h-1$ (and $F_{h-1}=F_h+e_h+\lis e_h$).

The integration in Eq.(\ref{3.ciao3}) can be performed by expanding in
series the exponential in the r.h.s. and
integrating term by term with respect to the gaussian integration
$P_{f_h,\lis C_{h-1}}(d\Psi^{(h)})$. This gives rise
to an expansion for $\WW_{\b,L}(A,\phi)$, which can be conveniently represented in terms of
{\it Gallavotti-Nicol\`o trees}, as described in Section 3.3 of \cite{GM} and in the next subsection.

\subsubsection{Tree expansion.}\label{b4}

For each $n\ge 0$ and $m\ge 2$, we introduce a
family $\TT_{h,n}^{m}$ of rooted labelled trees, defined in a way similar to
the family $\TT_{h,n}$ described in Section 3.3. of \cite{GM} (which we refer to
for more details), with the following minor modifications:
\begin{enumerate}
\item $\TT_{h,n}^{m}$ has $n+m$ endpoints (rather than $n$);
$n$ of them are called normal endpoints and $m$ of them are
called special endpoints; moreover, the special endpoints can be either of type $A$ or
of type $\phi$. If $v$ is a normal endpoint, then, as in \cite{GM}, it is associated to
one of the monomials
with four or more Grassmann fields contributing to $\RR {\cal
V}^{(0)}(\Psi^{(\le h_v-1)})$; if $v$ is a special endpoint of type $A$,
then it is associated to one of the monomials contributing to
$(A\Psi^{(\le h_v-1)},T_{h_v-1}\Psi^{(\le h_v-1)})-(A\Psi^{(\le h_v-2)},T_{h_v-1}\Psi^{(\le h_v-2)})$;
if $v$ is a special endpoint of type $\phi$,
then it is associated to one of the monomials contributing to
$(\phi, Q^{(h_v-1)}\Psi^{(h_v-1)})$.
\item Among the sets of field labels and external field labels associated to the vertex $v$,
denoted by $I_v$ and $P_v$ in \cite{GM}, we distinguish the field labels of type $A$, $\psi$
and $\phi$; we correspondingly introduce the sets $P_v^A$, $P_v^\psi$, $P_v^\phi$, $I_v^A$,
$I_v^\psi$ and $I_v^\phi$. All the trees contributing to $\WW_{\b,L}(A,\phi) + \b L^{2}F_{h_{\b}}$ 
are characterized by the fact that $P_{v_0}^\psi=\emptyset$ and $P_{v_0}^A\cup P_{v_0}^\phi\neq\emptyset$.
\end{enumerate}

Given $\t\in\TT^{m}_{h,n}$, the sets $P_v^\#$, $\#\in\{A,\phi,\psi\}$, satisfy several constraints
that depend on $\t$ (see \cite{GM}). In particular:
\begin{enumerate}
\item[a.] denoting by $v_0$ the vertex immediately
following the root on $\t$, we have $|P_{v_0}^\psi|=0$ and $|P_{v_0}^A|+|P_{v_0}^\phi|>0$;
if $v>v_0$, then $|P_v^\psi|>0$;
\item[b.] for any $v\in\t$, sets of external field labels such that $|P_v|=|P_v^\psi|=2$ are not allowed (this follows from the
definition of localization Eqs.(\ref{2.loc})-(\ref{2.30a}) and from the choice of ``reabsorbing" at each
step the local part into the gaussian integration, see Eqs.(\ref{2.31aba}), (\ref{2.35}), (\ref{3.ciao3}));
\item[c.] if $v$ is not an endpoint, then
sets of external field labels such that $|P_v|=2$ and $|P_v^\psi|=|P_v^\phi|=1$ or such that
$|P_v|=3$, $|P_v^A|=1$ and $|P_v^\psi|=2$  are not allowed (this follows from the
definition of localization Eqs.(\ref{2.locB})-(\ref{2.30aab}) and from the choice of ``reabsorbing" at
each step the bilinear terms $(\phi,\lis W^{(h)}_{2,\o}\Psi)$ into the definition of $(\phi, Q^{(h)}\Psi)$
and the ``vertex" terms $(A\Psi,W^{(h)}_{\o,\o'}\Psi)$ into the definition of $(A\Psi,T_{h-1}\Psi)$,
see Eq.(\ref{2.31azd})).
\end{enumerate}

As in \cite{GM}, we denote by ${\cal P}_\t$ the family of all the
choices of $P_v^\#$ compatible with these constraints and by ${\bf P}$ the elements of
${\cal P}_\t$. The generating functional can be expressed as a sum over trees in the following fashion (analogous to Eqs.(3.77),(3.79),(3.87),(3.88) of \cite{GM}):
\be \WW_{\b,L}(A,\phi) + \b L^{2}F_{h_{\b}} = \sum_{n\ge 0}\,\sum_{m\ge 2}\,
\sum_{h\ge h_\b}\,\sum_{\t\in\TT^{m}_{h,n}}\sum_{{\bf P}\in\PPP_\t}\sum_{T\in{\bf T}}
 \WW^{(h)}(\t,{\bf P},T)\;,\label{3.expn}\ee
where, as explained in \cite{GM}, ${\bf T}$ is a suitable family of {\it spanning trees}.
The contribution $\WW^{(h)}(\t,{\bf P},T)$ can be further rewritten as
\be \WW^{(h)}(\t,{\bf P},T)=\int d\xx_{v_0}\tilde A(P^A_{v_0})
 \,\tilde \phi(P^\phi_{v_0})\,W^{(h)}_{\t,{\bf P},T}(\xx_{v_0})\;,\label{3.expn1} \ee
where
\be \tilde A(P^A_{v_0})=\prod_{f\in P_{v_0}^A}e^{-i(\pp_F^{\o(f)}-\pp_F^{\o'(f)})\cdot\xx(f)}
A^{\o(f)-\o'(f)}_{\xx(f),\sharp(f)}\;,\qquad
\tilde \phi(P^\phi_{v_0})=\prod_{f\in P_{v_0}^\phi}e^{i\e(f)\pp_F^{\o(f)}\cdot\xx(f)}
\phi^{\e(f)}_{\xx(f),\s(f),\o(f)}\label{2.44}\ee
and, calling $v_i^*,\ldots,v_n^*$ the endpoints of $\t$, putting
$h_i=h_{v_i^*}$ and denoting by $K_{v_i^*}^{(h_i)} (\xx_{v_i^*})$ the kernels
associated to the endpoints,
\bea &&
W_{\t,\PP, T}(\xx_{v_0}) =\left[\prod_{i=1}^n
K_{v_i^*}^{(h_i)} (\xx_{v_i^*})\right] \;\cdot\label{2.50}\\ &&\cdot\;
\Bigg\{\prod_{v\ {\rm not}\ {\rm  e.p.}}\frac1{s_v!} \int
dP_{T_v}({\bf t}_v)\;{\rm det}\, G^{h_v,T_v}({\bf t}_v)\Biggl[
\prod_{l\in T_v} \d_{\o^-_l,\o^+_l} \d_{\s^-_l,\s^+_l}\,
\big[g^{(h_v)}_{\o_l}(\xx_l-\yy_l)\big]_{\r^-_l,\r^+_l}\,\Biggr]
\Bigg\}\;.\nn\eea
In the latter equation,  $dP_{T}({\bf t})$ is a probability measure with
support on a set of ${\bf t}$ such that $t_{ii'}={\bf u}_i\cdot{\bf u}_{i'}$
for some family of vectors ${\bf u}_i\in \RRR^s$ of unit norm. Finally
$G^{h,T}({\bf t})$ is a Gram matrix, whose elements
are given by, see \cite{GM}, Eq.(3.83):
\be G^{h,T}_{ij,i'j'}=t_{ii'}
\d_{\o^-_l,\o^+_l} \d_{\s^-_l,\s^+_l}\,
\big[g^{(h)}_{\o_l}(\xx_{ij}-\yy_{i'j'})\big]_{\r^-_l,\r^+_l}\;;
\label{2.48}\ee
this matrix takes into account all the possible contractions of fields not
involved in the spanning tree $T$. See \cite{GM} for more details.
The effective potential $\VV^{(h)}$ and the effective source term $B^{(h)}$ admit
representations very similar to Eqs.(\ref{3.expn})-(\ref{3.expn1})-(\ref{2.50}), the main difference
being that these are expressed as sums over trees and field labels satisfying slightly different
constraints: i.e., the trees contributing to $\VV^{(h)}$ do not have special endpoints and
are associated to external field labels such that
$P_{v_0}^A=P_{v_0}^\phi=\emptyset$ and $|P_{v_0}^\psi|>0$;
the trees contributing to $B^{(h)}$ have at least one special endpoint, 
and are associated to external field labels such that
$|P_{v_0}^\psi|>0$ and $|P_{v_0}^A|+|P_{v_0}^\phi|>0$.

\subsubsection{The kernels of the special endpoints of type $A$}

In order to prove Theorem 2 and Proposition 1, we will be particularly concerned with estimating
the contributions with $|P^A_{v_0}|=2$ and $|P^\phi_{v_0}|=0$ or with
$|P^A_{v_0}|=1$ and $|P^\phi_{v_0}|+|P_{v_0}^\psi|=2$. The key estimate that we preliminarily
need to prove is 
\be ||\hat W^{(k)}_{\sharp;\o,\o'}
(\kk',\pp') ||\le (\const.)|U|2^{\th k}\;,\label{8.1}\ee
for all $k\le 0$ and with $\th\in(0,1)$, uniformly in $\kk',\pp'$.
Note that Eq.(\ref{8.1}) implies, in particular, that the kernel of the
special endpoints of type $A$ is uniformly bounded as
\be ||T^{\o,\o'}_{\sharp,k}
(\kk',\pp') ||\le C_0\;,\label{8.2}\ee
for all $k\le 0$ and a suitable constant $C_0$, and that $|Z_{\m,h-1}-Z_{\m,h}|\le (\const.)|U|2^{\th h}$, as claimed after Eq.(\ref{2.zetamu}). Let us proceed by induction: we assume the validity of
Eq.(\ref{8.1}) for $k> h$ (so that Eq.(\ref{8.2}) is valid for all $k> h$) and prove it for $k=h$.
Using the tree expansion, we can rewrite:
\be \hat W^{(h)}_{\sharp;\o,\tilde\o}
(\kk',\pp')=\frac{1}{\b L^2}\sum_{n\ge 1}\,
\sum_{\t\in\TT^{1}_{h,n}}\sum_{{\bf P}\in\PPP_\t}^*\sum_{T\in{\bf T}}
\int\!\! d\xx_{v_0}\,e^{i(\pp_F^{\o}+\kk'+\pp')\xx-i(\pp_F^{\tilde\o}+\kk')\yy-i(\pp_F^\o-\pp_F^{\tilde\o}+\pp')\zz}W^{(h)}_{\t,{\bf P},T}(\xx_{v_0})\label{8.3}\ee
where the $*$ on the sum reminds that $P_{v_0}=P^A_{v_0}\cup P^\psi_{v_0}$,
$P^A_{v_0}=\{f_1\}$ and $P^\psi_{v_0}=\{f_2,f_3\}$ (with
$\o(f_1)=\o(f_2)=\o$, $\o'(f_1)=\o(f_3)=\tilde\o$, $\xx(f_1)=\zz$, $\xx(f_2)=\xx$, $\xx(f_3)=\yy$,
$\sharp(f_1)=\sharp$,
$\e(f_2)=-\e(f_3)=-$ and $\s(f_2)=\s(f_3)$). Using translation invariance, the representation Eq.(\ref{2.50}) and proceeding as in the proof of Theorem 2 of \cite{GM}, we get (see Eq.(3.93)
of \cite{GM})
\bea &&||\hat W^{(h)}_{\sharp;\o,\tilde\o}
(\kk',\pp')||\le  \sum_{n\ge 1}\sum_{\t\in {\cal T}^1_{h,n}}
\sum_{\PP\in{\cal P}_\t}^*\sum_{T\in{\bf T}}
\int\prod_{l\in T^*}
d(\xx_l-\yy_l) \left[\prod_{i=1}^n|K_{v_i^*}^{(h_i)}(\xx_{v_i^*})|\right]\cdot
\label{2.53}\\
&&\hskip3.truecm\cdot\Bigg[\prod_{v\ {\rm not}\ {\rm e.p.}}\frac{1}{s_v!}
\max_{{\bf t}_v}\big|{\rm det}\, G^{h_v,T_v}({\bf t}_v)\big|
\prod_{l\in T_v}
\big|\big|g^{(h_v)}_{\o_l}(\xx_l-\yy_l)\big|\big|\Bigg]\;.\nn\eea
The r.h.s. of this equation can be bounded dimensionally, using the scaling
properties of the propagators $g^{(h_v)}_{\o}(\xx)$ and of the Gram determinants
$\det G^{h_v, T_v}({\bf t}_v)$ (see Eqs.(3.92),(3.94) of \cite{GM}). Following the
proof of Theorem 2 of \cite{GM} and using in particular Eqs.(3.94)-(3.95)-(3.96), we get the
analogue of Eq.(3.97) of \cite{GM}, that is
\bea &&  ||\hat W^{(h)}_{\sharp;\o,\tilde\o}
(\kk',\pp')||\le \label{2.57}\\
&&\le \sum_{n\ge 1}\sum_{\t\in {\cal T}^1_{h,n}}
\sum_{\PP\in{\cal P}_\t}^*\sum_{T\in{\bf T}}
C^n \Big[\prod_{\substack{v\ {\rm not}\\ {\rm e.p.}}} \frac{1}{s_v!}
2^{{h_v}\left((\sum_{i=1}^{s_v}|P_{v_i}^\psi|)-|P_v^\psi|-3(s_v-1)\right)}\Big]
\Big[\prod_{v\ {\rm e.p.}} C^{\frac{|P_v^\psi|}2}
|U|^{\,\frac{|P_v^\psi|}2-1}\Big]\nn\eea
where $p_i=|P_{v_i^*}|$ and $C$ is a suitable positive constant, larger than the constant $C_0$
appearing in Eq.(\ref{8.2}) (in fact, in deriving this bound,
we estimated the kernel of the special endpoint of type $A$ by using the inductive hypothesis
Eq.(\ref{8.2})).
The r.h.s. of this expression can be rewritten in a convenient form, using
the analogues of Eqs.(3.98)-(3.100) of \cite{GM},
that is (recalling that $m$ is the number of special endpoints in $\t$ -- equal to 1 in the current
case),
\bea &&\sum_{\substack{h\ {\rm not}\\{\rm e.p.}}}h_v\Big[\big(\sum_{i=1}^{s_v}
|P_{v_i}^\psi|\big)-|P_v^\psi|\Big]=h(|I_{v_0}^\psi|-|P_{v_0}^\psi|)+\sum_{\substack{h\ {\rm not}\\
{\rm e.p.}}}(h_v-h_{v'})(|I_v^\psi|-|P_v^\psi|)\;,\nn\\
&&\sum_{\substack{h\ {\rm not}\\
{\rm e.p.}}}h_v(s_v-1)=h(n+m-1)+\sum_{\substack{h\ {\rm not}\\
{\rm e.p.}}}(h_v-h_{v'})(n(v)+m(v)-1)\;,\label{2.58}\eea
where: $v'$ is the vertex immediately preceding $v$ on $\t$; $I_v\supseteq P_v$
is the set of field labels associated to $v$ (i.e., including both the internal and the external fields
to $v$); $n(v)$ is the number of normal endpoints following $v$ on $\t$;
$m(v)$ is the number of special endpoints following $v$ on $\t$. Note that in the current case,
where there is only one special endpoint of type $A$, $m(v)=|P^A_v|$ and $m=|P^A_{v_0}|$.
Plugging Eq.(\ref{2.58}) into Eq.(\ref{2.57}) and using Eq.(3.100) of \cite{GM},
we get the analogue of Eq.(3.101) of \cite{GM}, that is,
\bea   ||\hat W^{(h)}_{\sharp;\o,\tilde\o}
(\kk',\pp')||&\le& \sum_{n\ge 1}\sum_{\t\in {\cal T}^1_{h,n}}
\sum_{\PP\in{\cal P}_\t}^*\sum_{T\in{\bf T}}
C^n  2^{h(3-|P_{v_0}^\psi|-|P_{v_0}^A|)}\cdot\Big[\prod_{\substack{v\ {\rm not}\\ {\rm e.p.}}}
\frac{1}{s_v!} 2^{(h_v-h_{v'})(3-|P_v^\psi|-|P_v^A|)}\Big]\cdot\nn\\
&&\cdot\Big[\prod_{v\ {\rm
e.p.}} 2^{h_{v'}(|P_v^\psi|+|P_v^A|-3)} \Big]\Big[\prod_{v\ {\rm
e.p.}} C^{\frac{|P_v^\psi|}2} |U|^{\,\frac{|P_v^\psi|}2-1}  \Big]\;,\label{2.61} \eea
where we used that, if $v$ is an endpoint, then $|I_v^\psi|=|P_v^\psi|$.
%
%
At this point, by Eqs.(3.102)-(3.103) of \cite{GM} and by the argument described after Eq.(3.103)
of \cite{GM}, we end up with the analogue of Eq.(3.104) of \cite{GM}, that is
\be  ||\hat W^{(h)}_{\sharp;\o,\tilde\o}
(\kk',\pp')||\le \g^{h(3-|P_{v_0}^\psi|-|P^A_{v_0}|+\th)}
\sum_{n\ge 1}C^n |U|^n\;.\label{2.61e}\ee
Recalling that in the current case $3-|P_{v_0}^\psi|-|P^A_{v_0}|=0$, this
proves the desired estimate on the kernels of the special endpoints of type $A$, Eq.(\ref{8.1}).
A similar strategy also allows us to prove that
\be  T^{\o,\o'}_{\sharp,h}(\kk',\pp') = T^{\o,\o'}_{\sharp,h}(\V0,\V0) + O\Big((|\kk'|+|\pp'|)+
|U|\,(|\kk'|+|\pp'|)^{\th}\Big)\;,\label{8.9}\ee
with $\th\in(0,1)$. Let us also recall that also the kernels $Q^{(h)}$ of the special endpoints of
type $\phi$ admit a uniform bound of the form $||Q^{(h)}||\le C_0$, see Eq.(3.114) of \cite{GM}.
We are now ready to give the proof of Theorem 2 and Proposition 1.

\subsubsection{The three-point function (Proof of Theorem 2).} \label{b5}

The goal is to bound $\hat G_{2,1;l}^{\b,L}(\kk,\pp)$ at $\kk\neq \pp_F^\pm$, $\pp\neq \V0$,
with $|\kk-\pp_F^\o|$ and $|\pp|$ sufficiently small (and $|\pp|\ll|\kk-\pp_F^\o|$),
for a given $\o\in\{+,-\}$. The three-point function, by definition, using
Eqs.(\ref{2.3e})-(\ref{3.g21j})-(\ref{3.g21}) and Eqs.(\ref{3.expn})-(\ref{2.50}), can be rewritten as
\be \hat G^{\b,L}_{2,1;l}(\kk,\pp)=\frac{e}{\b L^2}\sum_{j=1}^3(\vec\d_j)_l\h^j_{\vec p}
\sum_{n\ge 0}\,\sum_{h\ge h_\b}\,\sum_{\t\in\TT^{3}_{h,n}}\sum_{{\bf P}\in\PPP_\t}^{**}\sum_{T\in{\bf T}}
\int\!\! d\xx_{v_0}\,e^{i\kk(\xx-\yy)+i\pp(\xx-\zz)}W^{(h)}_{\t,{\bf P},T}(\xx_{v_0})\;,\label{8.6}\ee
where the $**$ on the sum over ${\bf P}$ reminds that
$P_{v_0}=P^A_{v_0}\cup P^\phi_{v_0}$,
$P^A_{v_0}=\{f_1\}$ and $P^\phi_{v_0}=\{f_2,f_3\}$, with
$\o(f_1)=\o'(f_1)=\o(f_2)=\o(f_3)=\o$, $\xx(f_1)=\zz$, $\xx(f_2)=\xx$, $\xx(f_3)=\yy$, $\sharp(f_1)=j$,
$\e(f_2)=-\e(f_3)=-$ and $\s(f_2)=\s(f_3)$. The reason why all the quasi-particle indices of the external legs are equal to $\o$ is that, by assumption, $|\pp|\ll|\kk-\pp_F^\o|\ll1$, so that
by momentum conservation all other choices of quasi-particle indices give zero contribution to $\hat G^{\b,L}_{2,1;l}(\kk,\pp)$. The trees contributing to $ \hat G^{\b,L}_{2,1;l}(\kk,\pp)$ have a few
more features that are worth remarking. First of all, among the three special endpoints of $\t\in\TT^3_{h,n}$, one of them is of type $A$ (let us call it $v_A$ and note that $f_1\in P^A_{v_A}$)
and the other two are of type $\phi$ (let us call them $v_\phi^+,v_\phi^-$, with
$f_2\in P^\phi_{v_\phi^-}$ and $f_3\in P^\phi_{v_\phi^+}$). Moreover,
let $h_{\kk}$ be the (negative) integer such that
$2^{h_{\kk}}\le |\kk-\pp_F^\o|<2^{h_{\kk}+1}$, and let $|\pp|$ be so small that
$2^{h_{\kk}-1}< |\kk+\pp-\pp_F^\o|<2^{h_{\kk}+2}$: then, only trees
with $|h_{v_\phi^\pm}-h_{\kk}|\le 1$ and with $h\le h_{\kk}+1$
contribute to $ \hat G^{\b,L}_{2,1;l}(\kk,\pp)$.\\

Now, let us distinguish in the r.h.s. of Eq.(\ref{8.6}), the contributions with $n=0$ and those
with $n\ge 1$. The latter can be bounded in a way completely analogous to
$\hat W^{(h)}_{\sharp;\o,\tilde\o}(\kk',\pp')$, with the following important differences.
In the current case $m=3$ and $m(v)=m^A(v)+m^\phi(v)$, with $m^A(v)=|P^A_v|$
(resp. $m^\phi(v)=|P^\phi_v|$)
the number of special endpoints of type $A$ (resp. type $\phi$)
following $v$ on $\t$. Taking this into account and following the same strategy used to bound
$\hat W^{(h)}_{\sharp;\o,\tilde\o}$, see Eq.(\ref{2.53})-(\ref{2.57})-(\ref{2.58}), we get
the analogue of Eq.(\ref{2.61}), that is we can bound the contributions with
$n\ge 1$ to the three point function by
\bea && e \sum_{n\ge 1}\sum_{h\le h_{\kk}+1}\sum_{\t\in {\cal T}^3_{h,n}}
\sum_{\PP\in{\cal P}_\t}^{**}\sum_{T\in{\bf T}}
C^n  2^{h(3-|P_{v_0}^\psi|-|P_{v_0}^A|-|P_{v_0}^\phi|)}\cdot\Big[\prod_{\substack{v\ {\rm not}\\ {\rm e.p.}}}\frac{1}{s_v!} 2^{(h_v-h_{v'})(3-|P_v^\psi|-|P_v^A|-|P_v^\phi|)}\Big]\cdot\nn\\
&&\hskip2.truecm\cdot\Big[\prod_{v\ {\rm
e.p.}} 2^{h_{v'}(|P_v^\psi|+|P_v^A|+|P_v^\phi|-3)} \Big]\Big[\prod_{v\ {\rm
e.p.}} C^{\frac{|P_v^\psi|}2} |U|^{\,\frac{|P_v^\psi|}2-1}  \Big]\;,\label{2.61az} \eea
Now note that, if $v$ is a special endpoint of type $\phi$, then $|P_v^\psi|+|P_v^A|+|P_v^\phi|-3=-1$ and that,   if $v$ is a special endpoint of type $A$, then $|P_v^\psi|+|P_v^A|+|P_v^\phi|-3=0$. Therefore, using the fact that $|P_{v_0}|=3$, we can rewrite Eq.(\ref{2.61az})
as
\bea && e\, 2^{-2 h_{\kk}}\sum_{n\ge 1}\sum_{h\le h_{\kk}+1}\sum_{\t\in {\cal T}^3_{h,n}}
\sum_{\PP\in{\cal P}_\t}^{**}\sum_{T\in{\bf T}}
C^n  \cdot\Big[\prod_{\substack{v\ {\rm not}\\ {\rm e.p.}}}\frac{1}{s_v!} 2^{(h_v-h_{v'})(3-|P_v|)}\Big]\cdot\nn\\
&&\hskip3.truecm\cdot\Big[\prod_{\substack{v\ {\rm normal}\\ {\rm
e.p.}}} 2^{h_{v'}(|P_v|-3)} \Big]\Big[\prod_{v\ {\rm
e.p.}} C^{\frac{|P_v^\psi|}2} |U|^{\,\frac{|P_v^\psi|}2-1}  \Big]\;.\label{2.61az8} \eea
The only potentially dangerous contributions to Eq.(\ref{2.61az8}) are those coming from a vertex
$v$ that is not an endpoint and such that $|P_v|=3$. By construction, such vertex has
necessarily $|P_v^A|=|P_v^\psi|=|P_v^\phi|=1$; on the other hand, by momentum conservation,
$0<h_v-h_{v'}\le 2$, simply because $|\pp|\ll|\kk-\pp_F^\o|\ll1$ and, therefore, the quasi-momenta associated to the external fields $\phi$ and $\Psi$ have essentially the same momentum scale
(i.e., the two scales differ at most by 1); in conclusion, the overall contribution coming from a vertex that is not an endpoint and such that $|P_v|=3$ can be bounded by an $O(1)$ constant
and gives no trouble. Therefore, Eq.(\ref{2.61az8}) implies that the overall contribution to $\hat G^{\b,L}_{2,1;l}(\kk,\pp)$
coming from trees with $n\ge 1$ can be bounded by $(\const.)|U|2^{(-2+\th) h_\kk}$. We are
left with the contributions coming from the trees with $n=0$, which read (defining
$\kk'=\kk-\pp_F^\o$)
\be e\,
\sum_{h,h'=h_\kk}^{h_{\kk}+1}Q^{(h)+}_{\kk'+\pp,\o}\,\hat g^{(h)}_{\o}(\kk'+\pp)
\Big[\sum_{j=1}^3(\vec\d_j)_l\h^j_{\vec p} T^{\o,\o}_{j,\lis h}(\kk',\pp)\Big]
g^{(h')}_{\o}(\kk')Q^{(h')-}_{\kk',\o}\;,\label{8.7}\ee
where $\lis h:=\max\{h,h'\}$. Using Eq.(\ref{8.1}), as well as the estimates on $Q^{(h)}$ and on
the two-point Schwinger function proved in \cite{GM}, see Eqs.(3.114),(3.120)--(3.122) of
\cite{GM}, we can rewrite Eq.(\ref{8.7}) as
\be e\,\Big(\hat S(\kk+\pp)
\big[\sum_{j=1}^3(\vec\d_j)_l\h^j_{\vec p} T^{\o,\o}_{j,h_\kk}(\kk',\pp)\big]
\hat S(\kk)\Big)\Big(1+O(|U|2^{\th h_\kk})\Big)\;,\label{8.8}\ee
for some $\th\in(0,1)$. Moreover, using Eq.(\ref{8.9}), we can further rewrite Eq.(\ref{8.8}) as
\be e\,\Big(\hat S(\kk+\pp)
Z_{l,h_\kk}\G_l(\vec p_F^{\,\o},\vec 0)
\hat S(\kk)\Big)\Big(1+O(2^{h_\kk})+O(|U|2^{\th h_\kk})\Big)\;,\label{8.10}\ee
where the vertex functions $Z_{l,h}$ were defined in Eq.(\ref{2.zetamu})
and we remind the reader that  $\G_1(\vec p_F^{\,\o},\vec 0)=-\s_2$ and
$\G_2(\vec p_F^{\,\o},\vec 0)=-\o\s_1$. Theorem 2 is an immediate corollary of the previous
estimates and, in particular, of Eqs.(\ref{2.61az8})--(\ref{8.10}). Using the fact that
$|Z_{l,h}-Z_{l,h-1}|\le (\const.)|U|2^{\th h}$, we also find that the constants $Z_{\m}$ in the
statement of Theorem 2 coincide with the infrared limit of the running coupling constants
$Z_{\m,h}$, i.e., $Z_\m=\lim_{h\to-\io}Z_{\m,h}$.
\qed

\subsubsection{The response function (Proof of Proposition 1).}

In order to prove Proposition 1, we start by deriving bounds on the current-current
response function $K^{\b,L}_{lm}(\xx-\yy)$ at distinct space-time points, $\xx\neq\yy$,
which can be expressed in terms of the tree expansion as follows:
\be K^{\b,L}_{lm}(\xx-\yy)=
e^2v_0^2\sum_{j,j'=1}^3(\vec\d_j)_l(\vec \d_{j'})_m
\sum_{n\ge 0}\,\sum_{h\ge h_\b}\,\sum_{\t\in\TT^{2}_{h,n}}\sum_{{\bf P}\in\PPP_\t}^{***}
\sum_{T\in{\bf T}}\int\!\! d\xx_{v_0}^*\,(H_{j,j'}*W^{(h)}_{\t,{\bf P},T})(\xx_{v_0})\;,\label{8.11}\ee
where the $***$ on the sum over ${\bf P}$ reminds that
$P_{v_0}=P^A_{v_0}=\{f_1,f_2\}$, with
$\o(f_1)-\o'(f_1)=\o(f_2)-\o'(f_2)=0$, $\xx(f_1)=\xx$, $\xx(f_2)=\yy$, $\xx(f_3)=\yy$, $\sharp(f_1)=j$
and $\sharp(f_2)=j'$; moreover, the $*$ over $d\xx_{v_0}$ reminds that
we are integrating over all the variables in $\xx_{v_0}$ {\it but} $\xx$ and $\yy$, which are fixed
(and distinct); finally, $H_{j,j'}$ is the Fourier transform of $\h^j_{\vec p}\h^{j'}_{-\vec p}$
and $(H_{j,j'}*W^{(h)}_{\t,{\bf P},T})$ denotes the convolutions between $H_{j,j'}$ and $W_{\t,{\bf P},T}$.
The trees contributing to $ K^{\b,L}_{lm}$ have two
special endpoints of type $A$; we call them $v_1$ and $v_2$, we denote by $v^*$ the
rightest vertex such that $v^*\le v_1,v_2$ with respect to the partial order induced by the tree
and by $h^*$ its scale.
Proceeding as in the previous subsection, we distinguish the contributions to $ K^{\b,L}_{lm}$
coming from trees of order $n=0$ or $n\ge 1$; we denote the two by $K^{(0)}_{lm}$ and $K^{(1)}_{lm}$, respectively. The latter can be bounded by
(using a notation analogous to the one used in Eq.(\ref{2.53}))
\bea &&||K^{(1)}_{lm}(\xx-\yy)||\le  e^2 v_0^2 \sum_{j,j'=1}^3\,||H_{j,j'}||_{L_1}
\sum_{n\ge 1}\sum_{h\ge h_\b}\sum_{\t\in {\cal T}^2_{h,n}}
\sum_{\PP\in{\cal P}_\t}^{***}\sum_{T\in{\bf T}}
\int\prod_{l\in T^*}^*d(\xx_l-\yy_l) \cdot\label{8.13}\\
&&\qquad\cdot\left[\prod_{i=1}^n|K_{v_i^*}^{(h_i)}(\xx_{v_i^*})|\right]\Bigg[\prod_{v\ {\rm not}\ {\rm e.p.}}\frac{1}{s_v!}
\max_{{\bf t}_v}\big|{\rm det}\, G^{h_v,T_v}({\bf t}_v)\big|
\prod_{l\in T_v}
\big|\big|g^{(h_v)}_{\o_l}(\xx_l-\yy_l)\big|\big|\Bigg]\;,\nn\eea
where $\prod^*_{l\in T^*}$ is the product over all lines of the (modified) spanning tree $T^*$,
but one line belonging to the subtree $T_{\xx,\yy}\subset T^*$ connecting $\xx$ with $\yy$
and contained into the cluster $v^*$ but not in any smaller one; let us call $\lis l$ this special line.
Eq.(\ref{8.13}) can be bounded in a way analogous to Eq.(\ref{2.53}), with the important difference
that the $L_1$ norm of the propagator associated to $\lis l$ is replaced by its $L_\io$
norm, which has a factor $2^{3 h^*}$ more as compared to the $L_1$ norm;
moreover, in order to take into account
the decay between $\xx$ and $\yy$, we can extract a factor $\frac{C_N^n}{1+
(2^{h^*}|\xx-\yy|)^{N}}$ from the product of the propagators in the spanning tree
(here $N\ge 1$ and $C_N$ is a suitable positive constant); we are still left with an expression
that can be estimated in the same way as Eq.(\ref{2.53}), thus leading to the upper bound
\bea &&||K^{(1)}_{lm}(\xx-\yy)||\le e^2v_0^2 \sum_{n\ge 1}\sum_{h\ge h_\b}\sum_{\t\in {\cal T}^2_{h,n}}
\sum_{\PP\in{\cal P}_\t}^{***}\sum_{T\in{\bf T}}
C^n  2^{h(3-|P_{v_0}|)}\cdot 2^{3 h^*}\cdot \frac{C_N^n}{1+(2^{h^*}|\xx-\yy|)^N}
\cdot\nn \\
&&\cdot\Big[\prod_{\substack{v\ {\rm not}\\ {\rm e.p.}}}\frac{1}{s_v!} 2^{(h_v-h_{v'})(3-|P_v|)}\Big]\cdot\Big[\prod_{v\ {\rm
e.p.}} 2^{h_{v'}(|P_v|-3)} \Big]\Big[\prod_{v\ {\rm
e.p.}} C^{\frac{|P_v^\psi|}2} |U|^{\,\frac{|P_v^\psi|}2-1}  \Big]\;.\label{8.12}\eea
Proceeding as in the previous subsections, we see that this can be further bounded
as
\be ||K^{(1)}_{lm}(\xx-\yy)||\le e^2 \sum_{n\ge 1}\sum_{h\ge h_\b}\sum_{h^*\ge h}
C^n |U|^n 2^{h(1+\th)} 2^{3 h^*}\frac{C_N^n}{1+(2^{h^*}|\xx-\yy|)^N}\;.
\label{8.14} \ee
Now, noting that, by construction, $h\ge h^*$ and exchanging the order of summation over $h$ and $h^*$, we get (picking $N=5$ and $|U|$ small enough)
\be ||K^{(1)}_{lm}(\xx-\yy)||\le (\const.) e^2 |U|\sum_{h^*\ge h_\b}  \frac{2^{h^*(4+\th)}}{1+(2^{h^*}|\xx-\yy|)^5}\;,\label{8.16}\ee
which implies
\be  ||K^{(1)}_{lm}(\xx-\yy)||\le (\const.) e^2 |U|\frac{1}{1+|\xx-\yy|^{4+\th}}\label{8.17}\ee
and, therefore, in the limit $\b,L\to\infty$ (which exists by the uniformity of the bounds and by the term by term convergence of the series, see Appendix D of \cite{GM} for more details),
the Fourier transform of $K^{(1)}$ is continuous and continuously differentiable
for all $\pp\in\RRR\times\BBB$ (in particular at $\pp=\V0$).\\

We are now left with the contributions to the response function coming from the trees with $n=0$,
which read
\bea && K^{(0)}_{lm}(\xx-\yy)=-2
e^2v_0^2\sum_{j,j'=1}^3(\vec\d_j)_l(\vec \d_{j'})_m
\sum_{h,h'\ge h_\b} \sum_{\o,\o'}\int\frac{d\kk' d\pp'}{(2\p)^2|\BBB|^2}
e^{i(\pp_F^\o-\pp_F^{\o'}+
\pp')(\xx-\yy)}\cdot\nn\\
&&\cdot\h^j_{\vec p_F^{\o}-\vec p_F^{\o'}+\vec p'}\,\h^{j'}_{-(\vec p_F^{\o}-\vec p_F^{\o'}+\vec p')} \Tr\{\hat g^{(h)}_{\o}(\kk'+\pp')T^{\o,\o'}_{j,\lis h}(\kk',\pp')
\hat g^{(h')}_{\o'}(\kk')T^{\o',\o}_{j',\lis h}(\kk'+\pp',-\pp')\}\;,\nn\eea
where $\lis h=\max\{h,h'\}$ and the factor 2 in the r.h.s. takes into account
the summation over the spin degrees of freedom. It is important to notice that
the integral in the latter expression can be rewritten as $e^{i(\pp_F^\o-\pp_F^{\o'})(\xx-\yy)} f^{\o,\o';j,j'}_{h,h'}(\xx-\yy)$, with $f^{\o,\o';j,j'}_{h,h'}(\xx-\yy)$ a function admitting
the dimensional bound:
\be |\dpr_\xx^n f^{\o,\o';j,j'}_{h,h'}(\xx)|\le c_N2^{2(h+h')}\frac{2^{n\cdot\max\{h,h'\}}}{1+(2^{\max\{h,h'\}}|\xx|)^N}\;,
\label{8.18}\ee
for all $N\ge 0$. Eq.(\ref{8.18}) implies that the contributions  to $K^{(0)}(\xx-\yy)$ with $\o\neq\o'$
can be rewritten as $e^{i(\pp_F^\o-\pp_F^{-\o})(\xx-\yy)} F_{\o,-\o}(\xx-\yy)$, with
\be |\dpr_\xx^n F_{\o,-\o}(\xx)|\le e^2 c_N\sum_{h,h'}2^{2(h+h')}\frac{2^{n\cdot\max\{h,h'\}}}{1+(2^{\max\{h,h'\}}|\xx|)^N}\le e^2\frac{c_N'}{1+|\xx|^{4+n}}\;,\label{8.19}\ee
for all $N\ge 4+n$. This implies, in particular, that we can rewrite
\bea && K^{(0)}_{lm}(\xx-\yy)=-2
e^2v_0^2\sum_{j,j'=1}^3(\vec\d_j)_l(\vec \d_{j'})_m
\sum_{h,h'\ge h_\b} \sum_{\o}\int\frac{d\kk' d\pp}{(2\p)^2|\BBB|^2}
e^{i\pp(\xx-\yy)}\h^j_{\vec p}\,\h^{j'}_{-\vec p'}\cdot\nn\\
&&\cdot \Tr\{\hat g^{(h)}_{\o}(\kk'+\pp)T^{\o,\o}_{j,\lis h}(\kk',\pp)
\hat g^{(h')}_{\o}(\kk')T^{\o,\o}_{j',\lis h}(\kk'+\pp,-\pp)\}+H^{(0)}_{lm}(\xx-\yy)\;,\nn\eea
where the Fourier transform of $H^{(0)}(\xx-\yy)$ is continuously differentiable in a neighborhood of $\pp=\V0$.
Using Eq.(\ref{8.9}) and the definition of $Z_{l,h}$,
we can further rewrite this expression as
\bea && K^{(0)}_{lm}(\xx-\yy)=-2
e^2v_0^2\sum_{h,h'\ge h_\b} \sum_{\o}Z_{l,h}Z_{m,h'}\int\frac{d\kk' d\pp}{(2\p)^2|\BBB|^2}
e^{i\pp(\xx-\yy)}\h^j_{\vec p}\,\h^{j'}_{-\vec p}\cdot\nn\\
&&\cdot \Tr\{\hat g^{(h)}_{\o}(\kk'+\pp)\G_l(\vec p_F^\o,\vec 0)
\hat g^{(h')}_{\o}(\kk')\G_m(\vec p_F^\o,\vec 0)\}+\lis H^{(0)}_{lm}(\xx-\yy)\;,\nn\eea
where the Fourier transform of $\lis H^{(0)}(\xx-\yy)$ is continuously differentiable in a
neighborhood of $\pp=\V0$. Finally, rewriting $Z_{l,h}=Z_l+O(|U|2^{\th h})$,
using the expression of the two-point function in terms of a sum of single scale propagators (see Eq.(3.121) of \cite{GM}) and using the definition of $\vec \G(\vec k,\vec p)$ in Eq.(\ref{1.cond2}),
we get
\bea K^{(0)}_{lm}(\xx-\yy)&=&-2
e^2v_0^2 Z_{l}Z_{m}\int\frac{d\kk d\pp}{(2\p)^2|\BBB|^2}
e^{i\pp(\xx-\yy)}\h^j_{\vec p}\,\h^{j'}_{-\vec p}
 \Tr\{\hat S(\kk+\pp)\G_l(\vec k,\vec 0)
\hat S(\kk)\G_m(\vec k,\vec 0)\}+\nn\\
&&+\widetilde H^{(0)}_{lm}(\xx-\yy)\;,\label{8.21}\eea
where  the Fourier transform of $\widetilde H^{(0)}(\xx-\yy)$ is continuously differentiable in a neighborhood of $\pp=\V0$.  Combining Eq.(\ref{8.21}) with the explicit expression Eq.(\ref{1.9})
for the two point function and with the bounds
derived above on $K^{(1)}$, we finally obtain the statement of Proposition 1 (the parity
property of $R(\pp)$ stated in item 3 of Proposition 1 easily follows from the symmetry properties
listed in Appendix \ref{secB} and \ref{secD2}). \qed

\appendix
\section{The conductivity of the non interacting system}\label{appA}
\setcounter{equation}{0}
\renewcommand{\theequation}{\ref{appA}.\arabic{equation}}

In this Appendix we prove Eq.(\ref{1.cond3}), with $\s_{ij}|_{U=0}$ given by Eq.(\ref{1.cond1}):
\be \s_{ij}\big|_{U=0}=\frac2{3\sqrt3}\frac{2e^2v_0^2}{\hslash}\lim_{p_0\to 0^+}\int\frac{d
k_0}{2\p} \int_{\BBB}\frac{d\vec
k}{|\BBB|}\Tr\Big\{\frac{S_0(\kk+(p_0,0))-S_0(\kk)}{p_0} \G_i(\vec k,\vec
0)S_0(\kk)\G_j(\vec k,\vec 0)\Big\}\;,\label{f1}\ee
where $|\BBB|=8\p^2/(3\sqrt3)$ and
\be \G_i(\vec k,\vec 0)=\begin{pmatrix}
0& a_i(\vec k)\\
a^*_i(\vec k)&0\end{pmatrix}\;,\label{A.2}\ee
with $a_1(\vec
k)=\frac{2i}{3}\big[1-e^{i\frac32k_1}\cos\big(\frac{\sqrt3}{2}k_2\big)\big]$ and $a_2(\vec
k)=\frac{2}{\sqrt3}e^{i\frac32k_1}\sin\big(\frac{\sqrt{3}}2 k_2\big)$.

Let $\e>0$ be a small number, independent of $p_0$, to be
eventually sent to zero. In the integral to be evaluated, we
distinguish between the region where $|\O(\vec k)|\ge \e$ and the
region where $|\O(\vec k)|\le \e$:
\bea  \s_{ij}\big|_{U=0}&=&\frac2{3\sqrt3}\frac{2e^2v_0^2}{\hslash}\lim_{\e\to 0}
\lim_{p_0\to 0^+}\int\frac{d
k_0}{2\p} \int_{\BBB}\frac{d\vec
k}{|\BBB|}\Tr\Big\{\frac{S_0(\kk+(p_0,0))-S_0(\kk)}{p_0} \G_i(\vec k,\vec
0)\cdot\nn\\
&&\cdot S_0(\kk)\G_j(\vec k,\vec 0)\Big\}\cdot\Big[\c(|\O(\vec k)|\ge \e)+\c(|\O(\vec
k)|\le \e)\Big] \;,\label{f2}\eea
The integral associated to the region $\c(|\O(\vec k)|\ge \e)$ is
uniformly convergent as $p_0\to 0^+$: therefore, we can exchange the integral with the
limit and check that the integral of the limit is zero (simply because the integrand is odd in $k_0$).
Next, in the integral associated to the region
$\c(|\O(\vec k)|\le \e)$ we rewrite the propagator as the
relativistic propagator plus a correction (similarly, we rewrite
$\G_i$ as its relativistic limit plus a correction). The
corrections are associated to absolutely convergent integrals,
uniformly in $p_0$ as $p_0\to 0^+$,
and one can easily check that their contribution after having
taken $\e\to 0$ is equal to zero. We are left with (after having
changed variables and having included a further factor 2 coming
from the summation over the two Fermi points):
\bea&& \s_{ij}\big|_{U=0}=\frac2{\p}\frac{e^2}{h}\lim_{\e\to 0} \lim_{p_0\to
0^+}\Biggl\{\frac1{p_0}\int\frac{d k_0}{2\p} \int_{|\vec k'|\le
\e}d\vec k'
\frac{1}{k_0^2+|\vec k'|^2} \Tr\Big\{\s_j\begin{pmatrix}ik_0& ik_1'-k_2'\\
-ik_1'-k_2'&ik_0\end{pmatrix}\s_i\cdot\nn\\
&&\cdot\Big[\frac1{(k_0+p_0)^2+|\vec k'|^2}
\begin{pmatrix}i(k_0+p_0)& ik_1'-k_2'\\
-ik_1'-k_2'&i(k_0+p_0)\end{pmatrix}-\frac1{k_0^2+|\vec k'|^2}
\begin{pmatrix}ik_0& ik_1'-k_2'\\
-ik_1'-k_2'&ik_0\end{pmatrix}\Big]\Big\}\Biggr\}\;,\nn\eea
where $\s_i$, $i=1,2$, are the first two Pauli matrices. Now, if $i\neq j$, the r.h.s. of
this equation is equal to
\be\pm \frac2{\p}\frac{e^2}{h}\lim_{\e\to 0} \lim_{p_0\to
0^+}\frac1{p_0}\int\frac{d k_0}{2\p} \int_{|\vec k'|\le
\e}d\vec k'
\frac{4k_1'k_2'}{k_0^2+|\vec k'|^2}\Big(\frac1{(k_0+p_0)^2+|\vec k'|^2}-\frac1{k_0^2+
|\vec k'|^2}\Big)\;,\label{f4z}\ee
which is zero by the symmetry under the exchange $\vec k'\otto-\vec k'$. If, on the contrary,
$i=j$, we get:
\bea  \s_{ii}\big|_{U=0}&=&\frac2{\p}\frac{e^2}{h}\lim_{\e\to 0} \lim_{p_0\to
0^+}\frac1{p_0}\int\frac{d k_0}{2\p} \int_{|\vec k'|\le
\e}d\vec k' \frac{1}{k_0^2+|\vec k'|^2}\Biggl[\Big(\frac{2k_0^2}{k_0^2+|\vec k'|^2}-
\frac{2k_0(k_0+p_0)}{(k_0+p_0)^2+|\vec k'|^2}\Big)+\nn\\
&&+(-1)^i[2(k_1')^2-2(k_2')^2]\Big(\frac{1}{k_0^2+|\vec k'|^2}-
\frac{1}{(k_0+p_0)^2+|\vec k'|^2}\Big)\Biggr]\;.\label{f4}\eea
Now, the terms in the integral proportional to
$2(k_1')^2-2(k_2')^2$ are zero by the symmetry under the exchange
$k_1'\otto k_2'$. Therefore, we are left with:
\be \s_{ii}\big|_{U=0}=\frac2{\p}\frac{e^2}{h}\lim_{\e\to 0} \lim_{p_0\to
0^+}\frac1{p_0}\int\frac{d k_0}{2\p}
\int_{|\vec k'|\le \e}d\vec k' \frac{1}{k_0^2+|\vec
k'|^2}\Big(\frac{2k_0^2}{k_0^2+|\vec k'|^2}-
\frac{2k_0(k_0+p_0)}{(k_0+p_0)^2+|\vec k'|^2}\Big)\;,\label{f5}\ee
that is
\be \s_{ii}\big|_{U=0}=8\frac{e^2}{h}\lim_{\e\to 0} \lim_{p_0\to
0^+}\frac1{p_0}\int_0^\e dk\cdot k\int\frac{d k_0}{2\p}
\Big(\frac{k_0^2}{(k_0^2+k^2)^2}-
\frac{k_0(k_0+p_0)}{\big[(k_0+p_0)^2+k^2\big]\cdot\big[k_0^2+k^2\big]}\Big)
\;.\label{f6}\ee
The integral in $k_0$ can be evaluated by residues to give:
\bea && \int\frac{d k_0}{2\p i} \Big[\frac{k_0^2}{(k_0^2+k^2)^2}-
\frac{k_0(k_0+p_0)}{\big[(k_0+p_0)^2+k^2\big]\cdot\big[k_0^2+k^2\big]}\Big]=
\dpr_{k_0}\Big[\frac{k_0^2}{(k_0+ik)^2}\Big]_{k_0=ik}-\nn\\
&&
\Big[\frac{k_0(k_0+p_0)}{\big[(k_0+p_0)^2+k^2\big]\cdot\big[k_0+ik\big]}
\Big]_{k_0=ik}-\Big[\frac{k_0(k_0+p_0)}{\big[k_0+p_0+ik\big]\cdot
\big[k_0^2+k^2\big]} \Big]_{k_0+p_0=ik}\;,\label{f7}\eea
that is
\bea && \int\frac{d k_0}{2\p} \Big[\frac{k_0^2}{(k_0^2+k^2)^2}-
\frac{k_0(k_0+p_0)}{\big[(k_0+p_0)^2+k^2\big]\cdot\big[k_0^2+k^2\big]}\Big]=
\frac{p_0^2}{4k(p_0^2+4k^2)}\;.\label{f8}\eea
Plugging (\ref{f8}) into (\ref{f6}) gives
\be \s_{ii}\big|_{U=0}=8\frac{e^2}{h}\lim_{\e\to 0} \lim_{p_0\to
0^+}\frac{p_0}{16}\int_0^\e dk \frac1{k^2+p_0^2/4}= \frac{e^2}{h} \lim_{\e\to 0}
\lim_{p_0\to 0^+}\arctan (2\e/p_0)=\frac{e^2}{h}\frac{\p}{2}\;,\label{8.23}\ee
which is the desired result.
%
%
\section{Symmetry transformations}
\label{secB} \setcounter{equation}{0}
\renewcommand{\theequation}{\ref{secB}.\arabic{equation}}
In the Appendix we collect some symmetry properties of the fermionic action,
i.e., some transformation of the fermionic fields and of the external sources
that leave separately invariant both the gaussian fermionic integration
$P(d\Psi)$ and the interactions $\VV(\Psi)$, $(\Psi,\phi)$, $(A,J)$.
These symmetries will be also preserved by the multiscale integration and,
therefore, they will allow us to exclude the presence of
possibly dangerous terms in the effective action at scale $h$, see
Section \ref{sec1.2b} and Appendix \ref{secD2}. In the following, we denote by
$\s_1,\s_2,\s_3$ the standard Pauli matrices and we use the follwoing
convention for the Fourier transform of the $A$ field:
$A_{\xx,\sharp}=(\b L^2)^{-1}\sum_{\pp\in \bar\BBB_{\b,L}}
e^{-i\pp\xx}\hat A_{\pp,\sharp}$, where $\bar\BBB_{\b,L}=
2\p\b^{-1}\ZZZ\times\BBB_L$ and $\sharp\in\{+,-,1,2,3\}$.
\begin{lemma}\label{lem2.4}
For any choice of $M,\b,L$, the fermionic Gaussian
integration $P(d\Psi)$, the interaction $\VV(\Psi)$ and the source terms
$(\Psi,\phi)$, $(A,J)$, defined in Eqs.(\ref{2.3})-(\ref{2.3c}),
are separately invariant under the following transformations (here
$\g\in\{ext,int\}$ and
$\hat\f_{\kk,\s, int}=\hat\Psi_{\kk,\s}$, while $\hat\f_{\kk,\s, ext}=
\hat\phi_{\kk,\s}$;
whenever this will not create ambiguities, we shall drop the labels $\s$ and
$\g$, i.e., we shall use $\hat \f_\kk^\pm$ as a shorthand for
$\hat\f^\pm_{\kk,\s,\g}$):
\begin{enumerate}
\item[(1)] \underline{Spin flip}:
$\hat\f^{\varepsilon}_{\kk,\s,\g}\leftrightarrow
\hat\f^{\varepsilon}_{\kk,-\s,\g}$;
\item[(2)] \underline{Global $U(1)$}:
$\hat\f^{\varepsilon}_{\kk,\s,\g}\rightarrow \eu^{ i
\varepsilon \a_{\s}}\hat\f^{\varepsilon}_{\kk,\s,\g}$, with
$\a_{\s}\in \RRR$ independent of $\kk$;
\item[(3)] \underline{Spin $SO(2)$}:
$\Biggl(\begin{matrix}
\hat\f^{\varepsilon}_{\kk,\uparrow,\cdot,\g} \\
\hat\f^{\varepsilon}_{\kk,\downarrow,\cdot,\g} \end{matrix}\Biggr)
\rightarrow e^{-i\th\s_2}\Biggl(\begin{matrix}
\hat\f^{\varepsilon}_{\kk,\uparrow,\cdot,\g} \\
\hat\f^{\varepsilon}_{\kk,\downarrow,\cdot,\g} \end{matrix}\Biggr)$, with
$\th\in \mathbb{T}=\RRR/2\p\ZZZ$ independent of $\kk$;
\item[(4)] \underline{Discrete rotations}:
$\hat \f_{\kk}^-\to
e^{-i\vec k(\vec \d_3
-\vec \d_1)\frac{\s_3}2}\hat \f_{T\kk}^-$,
$\hat\f_{\kk}^+\to
\hat\f_{T\kk}^+e^{i\vec k(\vec \d_3
-\vec \d_1)\frac{\s_3}2}$, $\hat A_{\pp,\pm}\to \hat A_{T\pp,\pm}
e^{\mp i\frac{\vec p}2(\vec\d_3-\vec\d_1)}$ and $\hat A_{\pp,j}\to
\hat A_{T\pp,j+1}e^{-i\frac{\vec p}{2}(\vec\d_3-\vec\d_1)}$, with
$T\kk=(k_0,e^{-i\frac{2\p}{3}\s_2}\vec k)$;
\item[(5)] \underline{Complex conjugation}:
$\hat\f^{\varepsilon}_{\kk}\rightarrow
\hat\f^{\varepsilon}_{-\kk}$, $\hat A_{\pp,\pm}\to
\hat A_{-\pp,\pm}$, $\hat A_{\pp,j}\to -\hat A_{-\pp,j}$ and
$c\rightarrow c^{*}$, where $c$ is generic
constant appearing in $P(d\Psi)$, in $\VV(\Psi)$ or in $(A,J)$;
\item[(6.a)] \underline{Horizontal reflections}:
$\hat\f^{-}_{\kk}\to \s_1\hat\f^-_{R_h\kk}$,
$\hat\f^{+}_{\kk}\to \hat\f^+_{R_h\kk}\s_1$,
$\hat A_{\pp,\pm}\to\hat A_{R_h\pp,\mp}$ and
$\hat A_{\pp,j}\to-\hat A_{R_h\pp,r_h j}e^{-i\vec p(\vec\d_j-\vec\d_1)}$,
with $R_h\kk=(k_0,-k_1,k_2)$ and $r_h1=1$, $r_h2=3$, $r_h3=2$;
\item[(6.b)] \underline{Vertical reflections}:
$\hat\f^{\varepsilon}_{\kk}\rightarrow
\hat\f^{\varepsilon}_{R_v\kk}$, $\hat A_{\pp,\pm}\to\hat A_{R_v\pp,\pm}$
and $\hat A_{\pp,j}\to\hat A_{R_v\pp,r_v j}$,
with $R_v\kk=(k_0,k_1,-k_2)$ and $r_v1=1$, $r_v2=3$, $r_v3=2$;
\item[(7)] \underline{Particle-hole}:
$\hat\f^{-}_{\kk}\to i\hat\f^{+,T}_{P\kk}$,
$\hat\f^{+}_{\kk}\to i\hat\f^{-,T}_{P\kk}$,
$\hat A_{\pp,\pm}\to\hat A_{-P\pp,\pm}$
and $\hat A_{\pp,j}\to-\hat A_{-P\pp, j}$,
with $P\kk=(k_0,-k_1,-k_2)$;
\item[(8)] \underline{Inversion}:
$\hat\Psi^{-}_{\kk,\s}\to -i\s_3\hat\Psi^-_{I\kk,\s}$,
$\hat\Psi^{+}_{\kk,\s}\to -i\hat\Psi^+_{I\kk,\s}\s_3$,
$\hat\phi^{-}_{\kk,\s}\to i\s_3\hat\phi^-_{I\kk,\s}$,
$\hat\phi^{+}_{\kk,\s}\to i\hat\phi^+_{I\kk,\s}\s_3$,
$\hat A_{\pp,\pm}\to-\hat A_{I\pp,\pm}$
and $\hat A_{\pp,j}\to\hat A_{I\pp, j}$,
with $I\kk=(-k_0,k_1,k_2)$.
\end{enumerate}
\end{lemma}
\noindent{\bf Proof.} The proof of the fact that $P(d\Psi)$,
and $\VV(\Psi)$ are separately invariant under the
transformations of the $\Psi$ fields has already been discussed
in Section 3.1 of \cite{GM}. The fact that $(\Psi,\phi)=(\b L^2)^{-1}
\sum_{\kk,\s} (\hat\Psi^+_{\kk,\s}\hat\phi^-_{\kk,\s}+\hat\phi^+_{\kk,\s}
\hat\Psi^-_{\kk,\s})$ is invariant is apparent from the definitions. Therefore,
here we are left with proving only
the invariance of the term $(A,J)$ under the transformations (4) to (8) of the
list above. In order to verify these symmetries, it is convenient to rewrite
the source term in Fourier space:
\bea (A,J) &=& \frac{e}{(\b L^2)^2}\sum_{\substack{
\pp\in \bar\BBB_{\b,L}\\ \kk\in\BBB^*_{\b,L}}}\sum_{\t=\pm}
\hat A_{\pp,\t}\hat\Psi^+_{\kk+\pp,\s}n_\t\hat\Psi^-_{\kk,\s}+\label{A1.1}\\
&+&\frac{ev_0}{(\b L^2)^2}\sum_{\substack{\pp\in\bar\BBB_{\b,L}\\
\kk\in\DD^*_{\b,L}}}\sum_{j=1,2,3}
\hat A_{\pp,j}\hat\Psi^+_{\kk+\pp,\s}\big(i\s_+e^{-i\kk(\dd_j-\dd_1)}
-i\s_-e^{i(\kk+\pp)(\dd_j-\dd_1)}\big)\hat\Psi^-_{\kk,\s}\;,\nn\eea
where it is implicit that the terms in the sums with $\kk+\pp\not\in
\BBB^*_{\b,L}$ should be put equal to zero.\\

\noindent{\it Symmetry (4).} The term %
\be (*):=\sum_{\kk,\pp,\t}
\hat A_{\pp,\t}\hat\Psi^+_{\kk+\pp,\s}n_\t\hat\Psi^-_{\kk,\s}\label{A1.0}\ee
in the first line of Eq.(\ref{A1.1}) is changed under (4) as:
\be (*)\to \sum_{\kk,\pp,\t}\hat A_{T\pp,\t}e^{-i\t\frac{\vec p}{2}
(\vec\d_3-\vec\d_1)}\,\hat\Psi^+_{T(\kk+\pp),\s}
\big[e^{i(\vec k+\vec p)(\vec\d_3-\vec\d_1)\frac{\s_3}2}
n_\t e^{-i\vec k(\vec\d_3-\vec\d_1)\frac{\s_3}2}\big]\hat\Psi^-_{\kk,\s}\;.
\label{A1.2}\ee
Using the definition of $n_\pm$, we find that
\be \big[e^{i(\vec k+\vec p)(\vec\d_3-\vec\d_1)\frac{\s_3}2}
n_\t e^{-i\vec k(\vec\d_3-\vec\d_1)\frac{\s_3}2}\big]=
e^{i\vec p(\vec\d_3-\vec\d_1)
\frac{\s_3}2}n_\t=e^{i\t\frac{\vec p}2(\vec\d_3-\vec\d_1)}n_\t
\;.\label{A1.3}\ee
Plugging this identity into Eq.(\ref{A1.2}) we see that $(*)$ is invariant
under (4). Similarly, the term
\be (**):=\sum_{\kk,\pp,j}
\hat A_{\pp,j}\hat\Psi^+_{\kk+\pp,\s}\big(i\s_+e^{-i\kk(\dd_j-\dd_1)}
-i\s_-e^{i(\kk+\pp)(\dd_j-\dd_1)}\big)\hat\Psi^-_{\kk,\s}\;,\label{A1.00}\ee
in the second line of Eq.(\ref{A1.1}) is changed under (4) as:
\bea (**)\to
\sum_{\kk,\pp,j}&&
\hat A_{T\pp,j+1}e^{-i\frac{\vec p}{2}(\vec\d_3-\vec\d_1)}\cdot\label{A1.003}\\
&&\cdot
\hat\Psi^+_{T(\kk+\pp),\s}\big[e^{i(\kk+\pp)(\dd_3-\dd_1)
\frac{\s_3}{2}}\big(i\s_+e^{-i\kk(\dd_j-\dd_1)}
-i\s_-e^{i(\kk+\pp)(\dd_j-\dd_1)}\big)e^{-i\kk(\dd_3-\dd_1)}\big]
\hat\Psi^-_{\kk,\s}\;.\nn\eea
Using the definition of $\s_\pm$, we find that $e^{i(\kk+\pp)(\dd_3-\dd_1)
\frac{\s_3}{2}}\s_\pm e^{-i\kk(\dd_3-\dd_1)}=
\s_\pm e^{\pm i(\kk+\frac{\pp}{2})(\dd_3-\dd_1)}$ and, therefore,
\bea && \big[e^{i(\kk+\pp)(\dd_3-\dd_1)
\frac{\s_3}{2}}\big(i\s_+e^{-i\kk(\dd_j-\dd_1)}
-i\s_-e^{i(\kk+\pp)(\dd_j-\dd_1)}\big)e^{-i\kk(\dd_3-\dd_1)}\big]=
\label{A1.004}\\&&\hskip6.truecm =
e^{i\frac{\pp}{2}(\dd_3-\dd_1)}\big(i\s_+ e^{-i\kk(\dd_j-\dd_3)}-i\s_-
e^{i(\kk+\pp)(\dd_j-\dd_3)}\big)\;.\nn\eea
Plugging this identity into Eq.(\ref{A1.003}) and using the fact that
$\kk(\dd_j-\dd_3)=(T\kk)(\dd_{j+1}-\dd_1)$ we see that also $(**)$ is invariant
under (4).\\

\noindent{\it Symmetry (5).}
The term $(*)$ is changed under (5) as:
\be (*)\to \sum_{\kk,\pp,\t}
\hat A_{-\pp,\t}\hat\Psi^+_{-(\kk+\pp),\s}n_\t\hat\Psi^-_{-\kk,\s}\;,
\label{A1.005}\ee
which is the same as $(*)$. Similarly, the term $(**)$ is changed under (5) as:
\be (**)\to\sum_{\kk,\pp,j}
(-\hat A_{-\pp,j})\hat\Psi^+_{-(\kk+\pp),\s}\big(-i\s_+e^{+i\kk(\dd_j-\dd_1)}
+i\s_-e^{-i(\kk+\pp)(\dd_j-\dd_1)}\big)\hat\Psi^-_{-\kk,\s}\;,
\label{A1.006}\ee
which is the same as $(**)$.\\

\noindent{\it Symmetry (6.a).}
The term $(*)$ is changed under (6.a) as:
\be (*)\to \sum_{\kk,\pp,\t}\hat A_{R_h\pp,-\t}
\hat\Psi^+_{R_h(\kk+\pp),\s}\s_1
n_\t\s_1\hat\Psi^-_{R_h\kk,\s}\;.\label{A1.007}\ee
Using the fact that $\s_1 n_\t \s_1=n_{-\t}$ we see that this term is
invariant under (5). The term $(**)$ is changed under (6.a) as:
\be (**)\to\sum_{\kk,\pp,j}
(-\hat A_{R_h\pp,r_hj})e^{-i\pp(\dd_j-\dd_1)}\,
\hat\Psi^+_{R_h(\kk+\pp),\s}\big[\s_1(i\s_+e^{-i\kk(\dd_j-\dd_1)}
-i\s_-e^{+i(\kk+\pp)(\dd_j-\dd_1)}\big)\s_1\big]\hat\Psi^-_{R_h\kk,\s}
\label{A1.008}\ee
where
\be \big[\s_1(i\s_+e^{-i\kk(\dd_j-\dd_1)}
-i\s_-e^{+i(\kk+\pp)(\dd_j-\dd_1)}\big)\s_1\big]=-e^{i\pp(\dd_j-\dd_1)}(i\s_+
e^{i\kk(\dd_j-\dd_1)}-i\s_-e^{-i(\kk+\pp)(\dd_j-\dd_1)})\;.\label{A1.009}\ee
Using this identity and the fact that $\kk(\dd_j-\dd_1)=-(R_h\kk)(\dd_{r_hj}-
\dd_1)$, we see that $(**)$ is invariant under (6.a).\\

\noindent{\it Symmetry (6.b).}
The term $(*)$ is changed under (6.b) as:
\be (*)\to \sum_{\kk,\pp,\t}\hat A_{R_v\pp,\t}
\hat\Psi^+_{R_v(\kk+\pp),\s}n_\t\hat\Psi^-_{R_v\kk,\s}\;,\label{A1.010}\ee
which is obviously the same as $(*)$. The term $(**)$ is changed under (6.b)
as:
\be (**)\to\sum_{\kk,\pp,j}
\hat A_{R_v\pp,r_vj}\,
\hat\Psi^+_{R_v(\kk+\pp),\s}(i\s_+e^{-i\kk(\dd_j-\dd_1)}
-i\s_-e^{+i(\kk+\pp)(\dd_j-\dd_1)}\big)\hat\Psi^-_{R_v\kk,\s}\;.
\label{A1.011}\ee
Using the fact that $\kk(\dd_j-\dd_1)=(R_v\kk)(\dd_{r_v}-\dd_1)$ we see that
also $(**)$ is invariant under (6.b).\\

\noindent{\it Symmetry (7).}
The term $(*)$ is changed under (7) as:
\be (*)\to -\sum_{\kk,\pp,\t}\hat A_{-P\pp,\t}
\hat\Psi^{-,T}_{P(\kk+\pp),\s}n_\t\hat\Psi^-_{P\kk,\s}=
\sum_{\kk,\pp,\t}\hat A_{-P\pp,\t}
\hat\Psi^{+}_{P\kk,\s}n_\t\hat\Psi^-_{P(\kk+\pp),\s}
\;,\label{A1.012}\ee
which is the same as $(*)$. The term $(**)$ is changed under (7) as:
\bea && (**)\to\sum_{\kk,\pp,j}
\hat A_{-P\pp,j}
\,\hat\Psi^{-,T}_{P(\kk+\pp),\s}(i\s_+e^{-i\kk(\dd_j-\dd_1)}
-i\s_-e^{+i(\kk+\pp)(\dd_j-\dd_1)}\big)\hat\Psi^{+,T}_{P\kk,\s}=\nn\\
&&=\sum_{\kk,\pp,j}
\hat A_{-P\pp,j}
\,\hat\Psi^{+}_{P\kk,\s}(i\s_+e^{i(\kk+\pp)(\dd_j-\dd_1)}
-i\s_-e^{-i\kk(\dd_j-\dd_1)}\big)\hat\Psi^{-}_{P(\kk+\pp),\s}\;.\label{A1.013}
\eea
Using the fact that $\kk(\dd_j-\dd_1)=-(P\kk)(\dd_{r_v}-\dd_1)$ we see that
also $(**)$ is invariant under (7).\\

\noindent{\it Symmetry (8).}
The term (*) is changed under (8) as:
\be (*)\to \sum_{\kk,\pp,\t}\hat A_{I\pp,\t}
\hat\Psi^+_{I(\kk+\pp),\s}\big[\s_3 n_\t\s_3\big]
\hat\Psi^-_{I\kk,\s}\;,\label{A1.014}\ee
which is the same as $(*)$. The term $(**)$ is changed under (8) as:
\be (**)\to-\sum_{\kk,\pp,j}
\hat A_{I\pp,j}\,
\hat\Psi^+_{I(\kk+\pp),\s}\big[\s_3(i\s_+e^{-i\kk(\dd_j-\dd_1)}
-i\s_-e^{+i(\kk+\pp)(\dd_j-\dd_1)}\big)\s_3\big]\hat\Psi^-_{I\kk,\s}\;.
\label{A1.015}\ee
Using the fact that $\s_3\s_\pm\s_3=-\s_\pm$ and that
$\kk(\dd_j-\dd_1)=(I\kk)(\dd_j-\dd_1)$ we see that
also $(**)$ is invariant under (8).\qed

\section{Symmetry properties of the kernels}\label{secD2}
\setcounter{equation}{0}
\renewcommand{\theequation}{\ref{secD2}.\arabic{equation}}

In this appendix we prove Eqs.(\ref{1.2.50a})--(\ref{1.2.51b}). We start by
studying the symmetries of the kernel quadratic in the external field $A$
and by proving Eqs.(\ref{1.2.50a})--(\ref{1.2.50c}). Next we investigate the
symmetries of the kernel quadratic in the fermionic fields $\Psi$ and linear
in the external field $A$ and prove Eqs.(\ref{1.2.51a})-(\ref{1.2.51b}).

{\it The $AA$ kernel.} We consider the term quadratic in the
external fields $A_{\sharp}$ in the r.h.s. of Eq.(\ref{2.2.16a}),
which has the form (neglecting the dependence on the label $M$,
defining $\hat W_{\sharp,\flat}(\pp):=
\hat W_{0,2;(\sharp,\flat)}(\pp,-\pp)$ and assuming, without loss of
generality, that $\hat W_{\sharp,\flat}(\pp)=\hat W_{\flat,\sharp}(-\pp)$):
\be \frac1{\b L^2}\sum_\pp\Big[
\sum_{\t,\t'=+,-}\hat A_{\pp,\t}
\hat W_{\t,\t'}(\pp)\hat A_{-\pp,\t'}+2\sum_{\substack{\t=\pm,\\ j=1,2,3}}
\hat A_{\pp,\t}
\hat W_{\t,j}(\pp)\hat A_{-\pp,j}+\sum_{j,j'=1,2,3}
\hat A_{\pp,j}\hat W_{j,j'}(\pp)\hat A_{-\pp,j'}\Big]\label{D.1}\ee
which must be invariant under the symmetry transformations listed in Appendix
\ref{secB}. Using
symmetries (4)--(8) of Appendix \ref{secB}, we find that:
\bea  \hat W_{\t,\t'}(\pp)&=&\hat W_{\t,\t'}(T\pp)e^{i\frac{\pp}2(\dd_3-\dd_1)
(\t-\t')}=\hat W_{\t,\t'}^*(-\pp)=\nn\\
&=&\hat W_{-\t,-\t'}(R_h\pp)=\hat W_{\t,\t'}(R_v\pp)=
\hat W_{\t,\t'}(I\pp)\;,\label{D.2}\\
 \hat W_{\t,j}(\pp)&=&\hat W_{\t,j+1}(T\pp)e^{i\frac{\pp}2(\dd_3-\dd_1)(\t-1)}=
-\hat W_{\t,j}^*(-\pp)=\nn\\
&=&-\hat W_{-\t,r_hj}(R_h\pp)e^{-i\pp(\dd_j-\dd_1)}=
\hat W_{\t,r_vj}(R_v\pp)=-\hat W_{\t,j}(I\pp)\;,\label{D.3}\\
\hat W_{j,j'}(\pp)&=&\hat W_{j+1,j'+1}(T\pp)=\hat W_{j,j'}^*(-\pp)=\nn\\
&=&
\hat W_{r_hj,r_hj'}(R_h\pp)e^{i\pp(\dd_j-\dd_{j'})}=\hat W_{r_vj,r_vj'}(R_v\pp)
=\hat W_{j,j'}(I\pp)\;.\label{D.4}\eea
Eqs.(\ref{D.2})--(\ref{D.4}) imply that one can define natural covariant
matrix elements as:
\bea&& \widetilde W_{\t,\t'}(\pp):=e^{i\frac{\pp}{2}\dd_1(\t-\t')}
\hat W_{\t,\t'}(\pp)\;,\label{D.5a}\\
&&\widetilde W_{\t,j}(\pp):=
e^{-i\frac{\pp}{2}(\dd_j-\dd_1)(\t-1)}
\hat W_{\t,j}(\pp)\;,\label{D.5b}\\
&&\widetilde W_{j,j'}(\pp):=
e^{-i\frac{\pp}{2}(\dd_j-\dd_{j'})}\hat W_{j,j'}(\pp)\;,\label{D.5c}\eea
which satisfy the following natural transformation rules:
\bea && \widetilde W_{\t,\t'}(\pp)=\widetilde W_{\t,\t'}(T\pp)=
\widetilde W_{\t,\t'}^*(-\pp)=\widetilde W_{-\t,-\t'}(R_h\pp)=
\widetilde W_{\t,\t'}(R_v\pp)=\widetilde W_{\t,\t'}(I\pp)\;,\label{D.55}\\
&& \widetilde W_{\t,j}(\pp)=\widetilde W_{\t,j+1}(T\pp)=
-\widetilde W_{\t,j}^*(-\pp)=-\widetilde W_{-\t,r_hj}(R_h\pp)=
\widetilde W_{\t,r_vj}(R_v\pp)=-\widetilde W_{\t,j}(I\pp)\;,\nn\\
&& \widetilde W_{j,j'}(\pp)=\widetilde W_{j+1,j'+1}(T\pp)=
\widetilde W_{j,j'}^*(-\pp)=\widetilde W_{r_hj,r_hj'}(R_h\pp)=\widetilde
W_{r_vj,r_vj'}(R_v\pp)=\widetilde W_{j,j'}(I\pp)\;.\nn\eea
At first order in $\pp$, defining $\widetilde W_{\sharp,\flat}(\V0)=:
a_{\sharp,\flat}$ and $\dpr_{p_\m}\widetilde W_{\sharp,\flat}(\V0)=:
b^\m_{\sharp,\flat}$, with $\sharp,\flat\in\{+,-,1,2,3\}$ and $\m\in\{0,1,2\}$,
from the first of Eq.(\ref{D.55}) we get $a_{\t,\t'}=(a_{\t,\t'})^*
=a_{-\t,-\t'}$ (i.e., $a_{\t,\t'}=a+a'\t\t'$, for some $a,a'\in\RRR$) and
$b^\m_{\t,\t'}=0$, $\forall\m\in\{0,1,2\}$, that is:
\be \widetilde W_{\t,\t'}(\pp)=a+a'\t\t'+O(\pp^2)\quad\Rightarrow\quad
\hat W_{\t,\t'}(\pp)=e^{-i\frac{\vec p}2\vec\d_1(\t-\t')}\big[a+a'\t\t'\big]+
O(\pp^2)\;,\quad a,a'\in\RRR\;,\label{D.56}\ee
which proves Eq.(\ref{1.2.50a}).
Similarly, from the second of Eq.(\ref{D.55}), we get $a_{\t,j}=0$,
$b^{l}_{\t,j}=0$ for $l=1,2$ and $b^0_{\t,j}=b\t$, for some $b\in\RRR$,
that is
\be \widetilde W_{\t,j}(\pp)=b\t p_0+O(\pp^2)\quad\Rightarrow\quad
\hat W_{\t,j}(\pp)=b\t p_0+O(\pp^2)\;,\quad b\in\RRR\;,\label{D.57}\ee
which proves Eq.(\ref{1.2.50b}).
Finally, from the third of Eq.(\ref{D.55}), we get that $a_{j,j'}=c\d_{j,j'}+
c'$, for some $c,c'\in\RRR$ and $b^\m_{j,j'}=0$, for all $\m\in\{0,1,2\}$
and $j,j'\in\{1,2,3\}$; that is,
\be \widetilde W_{j,j'}(\pp)=c\d_{j,j'}+c'+O(\pp^2)\quad\Rightarrow\quad
\hat W_{\t,j}(\pp)=e^{i\frac{\vec p}{2}(\vec\d_j-\vec\d_{j'})}
\big[c\d_{j,j'}+c'\big]+O(\pp^2)\;,\quad c,c'\in\RRR\;,\label{D.57a}\ee
which proves Eq.(\ref{1.2.50c}).\\

{\it The $A\psi\psi$ kernel.} We consider the term quadratic in
$\Psi^{(i.r.)}$ and linear in $A_{\sharp}$ in the r.h.s. of Eq.(\ref{2.2.16a}),
which has the form (neglecting the dependence on the label $M$ and
defining $\hat W_{\sharp}(\kk,\pp):=
\hat W_{2,1;\sharp}(\kk+\pp,\kk,\pp)$):
\be \frac1{(\b L^2)^2}\sum_{\kk,\pp,\s}\Big[\sum_{\t=\pm}\hat A_{\t,\pp}
\hat\Psi^{(i.r.)+}_{\kk+\pp,\s}\hat W_{\t}
(\kk,\pp)\hat\Psi^{(i.r.)-}_{\kk,\s}+
\sum_{j=1,2,3}\hat A_{j,\pp}
\hat\Psi^{(i.r.)+}_{\kk+\pp,\s}\hat W_{j}
(\kk,\pp)\hat\Psi^{(i.r.)-}_{\kk,\s}\Big]\;,\label{D.6}\ee
which must be invariant under the symmetry transformations listed in Appendix
\ref{secB}. Using
symmetries (4)--(8) of Appendix \ref{secB}, we find that:
\bea && \hat W_{\t}(\kk,\pp)=e^{i\t\frac{\pp}2(\dd_3-\dd_1)}
e^{-i(\kk+\pp)(\dd_3-\dd_1)\frac{\s_3}{2}}
\hat W_{\t}(T\kk,T\pp)e^{i\kk(\dd_3-\dd_1)\frac{\s_3}{2}}
=\hat W_{\t}^*(-\kk,-\pp)=\label{D.7}\\
&&\quad=\s_1\hat W_{-\t}(R_h\kk,R_h\pp)\s_1=\hat W_{\t}(R_v\kk,R_v\pp)=
\hat W_{\t}^T(P(\kk+\pp),-P\pp)=\s_3\hat W_{\t}(I\kk,I\pp)\s_3\;,
\nn\\
&&\hat W_{j}(\kk,\pp)=e^{i\frac{\pp}2(\dd_3-\dd_1)}
e^{-i(\kk+\pp)(\dd_3-\dd_1)\frac{\s_3}{2}}
\hat W_{j+1}(T\kk,T\pp)e^{i\kk(\dd_3-\dd_1)\frac{\s_3}{2}}=
-\hat W_{j}^*(-\kk,-\pp)=\nn\\
&&\quad=-e^{i\pp(\dd_j-\dd_1)}\s_1\hat W_{r_hj}(R_h\kk,R_h\pp)\s_1=
\hat W_{r_vj}(R_v\kk,R_v\pp)=\label{D.8}\\
&&\quad=-\hat W_j^T(P(\kk+\pp),-P\pp)=
-\s_3\hat W_{j}(I\kk,I\pp)\s_3\;.\nn\eea
At $\kk=\pp_F^\o$ and $\pp=\V0$ and defining $\hat W_\t(\pp_F^\o,\V0)=:
\sum_{\m=0}^3a^\m_{\o,\t}\s_\m$, with $\s_0=1$ and $\s_1,\s_2,\s_3$ the
standard Pauli matrices, the last identity in Eq.(\ref{D.7}) reads:
%
%
\be a^0_{\o,\t}+a^1_{\o,\t}\s_1+a^2_{\o,\t}\s_2+a^3_{\o,\t}\s_3=
a^0_{\o,\t}-a^1_{\o,\t}\s_1- a^2_{\o,\t}\s_2+a^3_{\o,\t}\s_3\;,
\label{D.9}\ee
which implies $a^1_{\o,\t}=a^2_{\o,\t}=0$; given this fact,
the second identity in Eq.(\ref{D.7}) implies that
$a^0_{\o,\t}=(a^0_{-\o,\t})^*$ and $a^3_{\o,\t}=(a^3_{-\o,\t})^*$;
the third identity implies that $a^{0}_{\o,\t}$ is even in $\t$, while
$a^{3}_{\o,\t}$ is odd in $\t$; the fourth and fifth identity imply that both
$a^{0}_{\o,\t}$ and $a^3_{\o,\t}$ are even in $\o$. In conclusion,
\be \hat W_\t(\pp_F^\o,\V0)=a^0+\t a^3 \s_3\;,\label{D.10}\ee
with $a^0$ and $a^3$ two real constants. This proves Eq.(\ref{1.2.51a}).

Similarly, defining  $\hat W_j(\pp_F^\o,\V0)=:
\sum_{\m=0}^3a^\m_{\o,j}\s_\m$, the last identity in Eq.(\ref{D.8})
reads $a^0_{\o,j}+a^1_{\o,j}\s_1+a^2_{\o,j}\s_2+a^3_{\o,j}\s_3=
-a^0_{\o,j}+a^1_{\o,j}\s_1+a^2_{\o,j}\s_2-a^3_{\o,j}\s_3$, which implies that
$a^0_{\o,j}=a^3_{\o,j}=0$; the first identity in Eq.(\ref{D.8}) reads:
\be a^1_{\o,j}\s_1+a^2_{\o,j}\s_2=
e^{-i\o\frac{2\p}{3}\frac{\s_3}{2}}(a^1_{\o,j+1}\s_1+a^2_{\o,j+1}\s_2)
e^{i\o\frac{2\p}{3}\frac{\s_3}{2}}\label{D.11}\ee
which implies  $\Biggl(\begin{matrix}
a^1_{\o,j}\\ a^2_{\o,j}\end{matrix}\Biggr)=e^{-i\o\frac{2\p}{3}\s_2}
\Biggl(\begin{matrix}a^1_{\o,j+1}\\ a^2_{\o,j+1}\end{matrix}\Biggr)$;
the second identity in Eq.(\ref{D.8}) implies that $a^{1}_{\o,j}=
-(a^{1}_{-\o,j})^*$ and $a^{2}_{\o,j}=(a^{2}_{-\o,j})^*$; the third identity
implies that $a^1_{\o,j}=-a^1_{\o,r_hj}$ and $a^2_{\o,j}=a^2_{\o,r_hj}$;
the fourth identity implies that $a^{l}_{\o,j}=a^{l}_{-\o,r_vj}$, with $l=1,2$;
the fifth identity implies that $a^1_{\o,j}=-a^1_{-\o,j}$ and $a^2_{\o,j}=
a^2_{-\o,j}$. Using the second, third, fourth and fifth identities for $j=1$
immediately gives $a^1_{\o,1}=0$ and $a^2_{\o,1}=a\in\RRR$. At this point,
using the first identity, we get $a^1_{\o,2}=-a^1_{\o,3}=\o a\frac{\sqrt3}{2}$
and $a^2_{\o,2}=a^2_{\o,3}=-\frac{a}{2}$, which means
\be\hat W_1(\pp_F^\o)=a\s_2\;,\qquad \hat W_2(\pp_F^\o)=
a(\o\frac{\sqrt3}2\s_1-\frac12\s_2)\;,\qquad \hat W_3(\pp_F^\o)=
a(-\o\frac{\sqrt3}2\s_1-\frac12\s_2)\;,\label{D.12}\ee
that is
\be \hat W_j(\pp_F^\o,\V0)=ae^{i\o\frac{2\p}{3}(j-1)\s_3}\s_2\;,\label{D.13}\ee
which proves Eq.(\ref{1.2.51b}).\qed

\\

\vskip.05truecm {\bf Acknowledgements.} A.G. and V.M. gratefully acknowledge
financial support from the ERC Starting Grant CoMBoS-239694. We thank D.
Haldane for valuable discussions on the role of exact lattice Ward
Identities.

\end{document}